\documentclass[prd,twocolumn,showpacs,amsmath,amssymb,superscriptaddress,groupedaddress]{revtex4}
\usepackage{graphicx}
\usepackage{dcolumn}
\usepackage{bm}
\usepackage{multirow} 
\usepackage{subfigure}
\usepackage{amsmath}
\usepackage{lineno}
\usepackage{array}
\usepackage{rotating}

\newcommand{\ts}    {\ensuremath{\thinspace}}

\begin{document}

\hspace{5.2in} \mbox{Fermilab-Pub-12/488-E}

\title{Measurement of the semileptonic charge asymmetry in ${\bm B^0}$ meson mixing with the D0 detector}

%
\affiliation{LAFEX, Centro Brasileiro de Pesquisas F\'{i}sicas, Rio de Janeiro, Brazil}
\affiliation{Universidade do Estado do Rio de Janeiro, Rio de Janeiro, Brazil}
\affiliation{Universidade Federal do ABC, Santo Andr\'e, Brazil}
\affiliation{University of Science and Technology of China, Hefei, People's Republic of China}
\affiliation{Universidad de los Andes, Bogot\'a, Colombia}
\affiliation{Charles University, Faculty of Mathematics and Physics, Center for Particle Physics, Prague, Czech Republic}
\affiliation{Czech Technical University in Prague, Prague, Czech Republic}
\affiliation{Center for Particle Physics, Institute of Physics, Academy of Sciences of the Czech Republic, Prague, Czech Republic}
\affiliation{Universidad San Francisco de Quito, Quito, Ecuador}
\affiliation{LPC, Universit\'e Blaise Pascal, CNRS/IN2P3, Clermont, France}
\affiliation{LPSC, Universit\'e Joseph Fourier Grenoble 1, CNRS/IN2P3, Institut National Polytechnique de Grenoble, Grenoble, France}
\affiliation{CPPM, Aix-Marseille Universit\'e, CNRS/IN2P3, Marseille, France}
\affiliation{LAL, Universit\'e Paris-Sud, CNRS/IN2P3, Orsay, France}
\affiliation{LPNHE, Universit\'es Paris VI and VII, CNRS/IN2P3, Paris, France}
\affiliation{CEA, Irfu, SPP, Saclay, France}
\affiliation{IPHC, Universit\'e de Strasbourg, CNRS/IN2P3, Strasbourg, France}
\affiliation{IPNL, Universit\'e Lyon 1, CNRS/IN2P3, Villeurbanne, France and Universit\'e de Lyon, Lyon, France}
\affiliation{III. Physikalisches Institut A, RWTH Aachen University, Aachen, Germany}
\affiliation{Physikalisches Institut, Universit\"at Freiburg, Freiburg, Germany}
\affiliation{II. Physikalisches Institut, Georg-August-Universit\"at G\"ottingen, G\"ottingen, Germany}
\affiliation{Institut f\"ur Physik, Universit\"at Mainz, Mainz, Germany}
\affiliation{Ludwig-Maximilians-Universit\"at M\"unchen, M\"unchen, Germany}
\affiliation{Fachbereich Physik, Bergische Universit\"at Wuppertal, Wuppertal, Germany}
\affiliation{Panjab University, Chandigarh, India}
\affiliation{Delhi University, Delhi, India}
\affiliation{Tata Institute of Fundamental Research, Mumbai, India}
\affiliation{University College Dublin, Dublin, Ireland}
\affiliation{Korea Detector Laboratory, Korea University, Seoul, Korea}
\affiliation{CINVESTAV, Mexico City, Mexico}
\affiliation{Nikhef, Science Park, Amsterdam, the Netherlands}
\affiliation{Radboud University Nijmegen, Nijmegen, the Netherlands}
\affiliation{Joint Institute for Nuclear Research, Dubna, Russia}
\affiliation{Institute for Theoretical and Experimental Physics, Moscow, Russia}
\affiliation{Moscow State University, Moscow, Russia}
\affiliation{Institute for High Energy Physics, Protvino, Russia}
\affiliation{Petersburg Nuclear Physics Institute, St. Petersburg, Russia}
\affiliation{Instituci\'{o} Catalana de Recerca i Estudis Avan\c{c}ats (ICREA) and Institut de F\'{i}sica d'Altes Energies (IFAE), Barcelona, Spain}
\affiliation{Uppsala University, Uppsala, Sweden}
\affiliation{Lancaster University, Lancaster LA1 4YB, United Kingdom}
\affiliation{Imperial College London, London SW7 2AZ, United Kingdom}
\affiliation{The University of Manchester, Manchester M13 9PL, United Kingdom}
\affiliation{University of Arizona, Tucson, Arizona 85721, USA}
\affiliation{University of California Riverside, Riverside, California 92521, USA}
\affiliation{Florida State University, Tallahassee, Florida 32306, USA}
\affiliation{Fermi National Accelerator Laboratory, Batavia, Illinois 60510, USA}
\affiliation{University of Illinois at Chicago, Chicago, Illinois 60607, USA}
\affiliation{Northern Illinois University, DeKalb, Illinois 60115, USA}
\affiliation{Northwestern University, Evanston, Illinois 60208, USA}
\affiliation{Indiana University, Bloomington, Indiana 47405, USA}
\affiliation{Purdue University Calumet, Hammond, Indiana 46323, USA}
\affiliation{University of Notre Dame, Notre Dame, Indiana 46556, USA}
\affiliation{Iowa State University, Ames, Iowa 50011, USA}
\affiliation{University of Kansas, Lawrence, Kansas 66045, USA}
\affiliation{Kansas State University, Manhattan, Kansas 66506, USA}
\affiliation{Louisiana Tech University, Ruston, Louisiana 71272, USA}
\affiliation{Northeastern University, Boston, Massachusetts 02115, USA}
\affiliation{University of Michigan, Ann Arbor, Michigan 48109, USA}
\affiliation{Michigan State University, East Lansing, Michigan 48824, USA}
\affiliation{University of Mississippi, University, Mississippi 38677, USA}
\affiliation{University of Nebraska, Lincoln, Nebraska 68588, USA}
\affiliation{Rutgers University, Piscataway, New Jersey 08855, USA}
\affiliation{Princeton University, Princeton, New Jersey 08544, USA}
\affiliation{State University of New York, Buffalo, New York 14260, USA}
\affiliation{University of Rochester, Rochester, New York 14627, USA}
\affiliation{State University of New York, Stony Brook, New York 11794, USA}
\affiliation{Brookhaven National Laboratory, Upton, New York 11973, USA}
\affiliation{Langston University, Langston, Oklahoma 73050, USA}
\affiliation{University of Oklahoma, Norman, Oklahoma 73019, USA}
\affiliation{Oklahoma State University, Stillwater, Oklahoma 74078, USA}
\affiliation{Brown University, Providence, Rhode Island 02912, USA}
\affiliation{University of Texas, Arlington, Texas 76019, USA}
\affiliation{Southern Methodist University, Dallas, Texas 75275, USA}
\affiliation{Rice University, Houston, Texas 77005, USA}
\affiliation{University of Virginia, Charlottesville, Virginia 22904, USA}
\affiliation{University of Washington, Seattle, Washington 98195, USA}
\author{V.M.~Abazov} \affiliation{Joint Institute for Nuclear Research, Dubna, Russia}
\author{B.~Abbott} \affiliation{University of Oklahoma, Norman, Oklahoma 73019, USA}
\author{B.S.~Acharya} \affiliation{Tata Institute of Fundamental Research, Mumbai, India}
\author{M.~Adams} \affiliation{University of Illinois at Chicago, Chicago, Illinois 60607, USA}
\author{T.~Adams} \affiliation{Florida State University, Tallahassee, Florida 32306, USA}
\author{G.D.~Alexeev} \affiliation{Joint Institute for Nuclear Research, Dubna, Russia}
\author{G.~Alkhazov} \affiliation{Petersburg Nuclear Physics Institute, St. Petersburg, Russia}
\author{A.~Alton$^{a}$} \affiliation{University of Michigan, Ann Arbor, Michigan 48109, USA}
\author{A.~Askew} \affiliation{Florida State University, Tallahassee, Florida 32306, USA}
\author{S.~Atkins} \affiliation{Louisiana Tech University, Ruston, Louisiana 71272, USA}
% ...
%
%
\collaboration{The D0 Collaboration\footnote{with visitors from
$^{a}$Augustana College, Sioux Falls, SD, USA,
$^{b}$The University of Liverpool, Liverpool, UK,
$^{c}$UPIITA-IPN, Mexico City, Mexico,
$^{d}$DESY, Hamburg, Germany,
$^{e}$SLAC, Menlo Park, CA, USA,
$^{f}$University College London, London, UK,
$^{g}$Centro de Investigacion en Computacion - IPN, Mexico City, Mexico,
$^{h}$ECFM, Universidad Autonoma de Sinaloa, Culiac\'an, Mexico
and
$^{i}$Universidade Estadual Paulista, S\~ao Paulo, Brazil.
}} \noaffiliation
\vskip 0.25cm

\date{August 28, 2012}

\begin{abstract}
We present a measurement of the semileptonic mixing asymmetry for $B^0$ mesons, $a^d_{\text{sl}}$, 
using two independent decay channels: $B^0 \to \mu^+D^-X$, with $D^- \to K^+\pi^-\pi^-$; 
and $B^0 \to \mu^+D^{*-}X$, with $D^{*-} \to \bar{D}^0\pi^-$, $\bar{D}^0 \to K^+\pi^-$ (and charge conjugate processes). 
We use a data sample corresponding to $10.4$~fb$^{-1}$ of $p\bar{p}$ collisions at $\sqrt{s} = 1.96$~TeV, 
collected with the D0 experiment at the Fermilab Tevatron collider.
We extract the charge asymmetries in these two channels as a function of the visible proper 
decay length (VPDL) of the $B^0$ meson, correct for detector-related asymmetries using data-driven 
methods, and account for dilution from charge-symmetric processes using Monte Carlo simulation.
The final measurement combines four signal VPDL regions for each channel, yielding 
$a^d_{\text{sl}} = [0.68 \pm 0.45 \text{ (stat.)} \pm 0.14 \text{ (syst.)}]\%$. 
This is the single most precise measurement of this parameter, 
with uncertainties smaller than the current world average of $B$ factory measurements.
\end{abstract}

\maketitle


\section{\label{sec:intro}Introduction}

Fundamental asymmetries in the interactions of elementary particles
influence the large-scale behavior of the universe.
Of particular interest is the process of baryogenesis, 
whereby an initially symmetric system of particles and antiparticles produced by the Big Bang 
evolved into the observed matter-dominated universe of the present day. 
Current theoretical models, building on the work of Sakharov~\cite{sak}, 
require CP-symmetry violating processes in order for baryogenesis to have occurred 
in the very early universe~\cite{theo1,theo2, npmixing1, npmixing2}. 
As such, studies of asymmetries in particle physics experiments have an 
influence far beyond the scale that they probe directly.

CP symmetry implies that physical processes are invariant under the combined 
parity and charge conjugation transformations.  
The standard model (SM) of particle physics is not CP symmetric as it stands, 
due to a complex phase in the quark mixing matrix of the weak interaction,
which has been measured to be non-zero~\cite{pdg}. 
While such  SM processes introduce some degree of CP violation (CPV), 
the effects in the quark sector are far too weak to explain the observed matter dominance 
of the universe~\cite{huet}. 
Consequently, it is important to search for further non-SM sources of CPV. 

Studies of neutral $B$ meson oscillations, whereby a neutral meson changes into 
its own antiparticle via a box-diagram-mediated weak interaction~\cite{pdg}, can provide a sensitive probe for such CPV processes. 
The semileptonic mixing asymmetry, defined as:
\begin{flalign}
a^q_{\text{sl}} =  \frac{\Gamma(\bar{B}^0_q \rightarrow B^0_q \rightarrow \ell^+ X) - \Gamma(B^0_q \rightarrow \bar{B}^0_q \rightarrow \ell^- X)}
        {\Gamma(\bar{B}^0_q \rightarrow B^0_q \rightarrow  \ell^+ X) + \Gamma(B^0_q \rightarrow \bar{B}^0_q \rightarrow  \ell^- X)}, \label{eq:aqsl_def}
\end{flalign}
allows the effects of any CP-violating processes to be directly observed 
in terms of the resulting asymmetry of the decay products. 
Here $\ell$ denotes a charged lepton of any flavor, and
$q$ represents the flavor of the non-$b$ valence quark of the meson.

In the standard model, the semileptonic mixing asymmetry is related to the properties of the corresponding $B$ meson system, 
namely the mass difference $\Delta M_q = M(B^0_{qH}) - M(B^0_{qL})$, 
the decay-width difference $\Delta\Gamma_q = \Gamma(B^0_{qL}) - \Gamma(B^0_{qH})$, 
and the CP-violating phase $\phi_q$, by:
\begin{eqnarray}
a^q_{\text{sl}} = \frac{|\Gamma^q_{12}|}{|M^q_{12}|} \sin\phi_q = \frac{\Delta\Gamma_q}{\Delta M_q} \text{tan}\phi_q.
\end{eqnarray}
Here the states $B^0_{qH}$ and $B^0_{qL}$ are the heavy and light mass eigenstates 
of the $B$ meson system, which differ from the flavor eigenstates. $M^q_{12}$ and $\Gamma^q_{12}$ are 
respectively the off-diagonal elements of the mass and decay matrices~\cite{pdg}.

The standard model predictions~\cite{nierste} for both $a^s_{\text{sl}}$ and $a^d_{\text{sl}}$ are very small: 
\begin{eqnarray}
a^d_{\text{sl}} &=& (-0.041 \pm 0.006) \%, \label{eq:adsl_SM} \\
a^s_{\text{sl}} &=& (0.0019 \pm 0.0003) \%.
\end{eqnarray}
These predictions are effectively negligible compared to the current experimental precision.
Hence, the measurement of any significant deviation from zero is an unambiguous signal of new physics,
which could lead to order-of-magnitude enhancements of $|a^d_{\text{sl}}|$~\cite{anatomy}.

The $B^0$ semileptonic mixing asymmetry, $a^d_{\text{sl}}$, has been extensively studied by the $B$ factories operating at the 
$\Upsilon(4S)$ resonance, including measurements by the CLEO~\cite{cleo115,cleo116},
{\sc BaBar}~\cite{babar121, babar123}, and Belle~\cite{belle125} collaborations.
The current world average of these measurements is~\cite{pdg}:
\begin{eqnarray}
a^d_{\text{sl}} &=& (-0.05 \pm 0.56) \%. \label{eq:adsl_pdg}
\end{eqnarray}
Additional inclusive measurements from LEP~\cite{adsl_lep1, adsl_lep2, adsl_lep3} 
and D0~\cite{adsl_d0} are subject to contamination from $B_s^0$ mesons,
and the extraction of $a^d_{\text{sl}}$ relies upon assumptions about the contribution from $a^s_{\text{sl}}$.

The recent evidence for a non-zero dimuon charge asymmetry by the D0 experiment is 
sensitive to the linear combination of $B^0$ and $B^0_s$ mixing asymmetries, with approximately 
equal contributions from each source~\cite{d0_dimuon}. 
The measurement constrains a band in the $(a^d_{\text{sl}},a^s_{\text{sl}})$ plane, 
which is inconsistent with the SM prediction at the 3.9 standard deviations level. 
By dividing the sample into two components with different relative contributions from $B^0$ and $B^0_s$, 
the semileptonic asymmetries are measured to be:
\begin{eqnarray}
a^d_{\text{sl}}(\mu\mu) &=& (-0.12 \pm 0.52) \%, \label{eq:adsl_mumu} \\
a^s_{\text{sl}}(\mu\mu) &=& (-1.81 \pm 1.06) \%,
\end{eqnarray}
where the measurements have a correlation coefficient of $-0.799$.
The above extraction assumes that any new source of CPV entering the dimuon asymmetry 
does so through $B$ mixing. 
Alternative hypotheses, for example new sources of dimuons from non-SM processes, cannot be excluded.
 
Recent searches for CPV in $B_s^0 \to J/\psi\phi$ decays from the D0~\cite{jpsiphi_d0}, 
CDF~\cite{jpsiphi_cdf}, and LHCb~\cite{jpsiphi_lhcb} collaborations find agreement of the CP-violating 
phase $\phi_s$ with SM predictions.
Given the current body of experimental evidence, improved measurements of both $a^d_{\text{sl}}$ and $a^s_{\text{sl}}$ 
are required in order to constrain the possible sources of new physics in $B$ meson mixing and decay~\cite{uli_new}.

This article describes the measurement of the semileptonic mixing asymmetry for $B^0_d$ mesons, 
\begin{flalign}
&a^d_{\text{sl}} =  &\label{eq:adsl_def} \\
&\frac{\Gamma(\bar{B}^0 \rightarrow B^0 \rightarrow \ell^+ D^{(*)-}X) - \Gamma(B^0 \rightarrow \bar{B}^0 \rightarrow \ell^- D^{(*)+}X)}
        {\Gamma(\bar{B}^0 \rightarrow B^0 \rightarrow  \ell^+ D^{(*)-}X) + \Gamma(B^0 \rightarrow \bar{B}^0 \rightarrow  \ell^- D^{(*)+}X)},& 
\nonumber
\end{flalign}
without the use of initial-state flavor tagging. The flavor of the $B^0$ meson at the time of decay is determined
 by the charge of the muon in the semileptonic decay.
Two separate decay channels are used:
\begin{enumerate}
\item $B^0 \rightarrow \mu^+ \nu D^{-} X$, 
\\with $D^{-} \rightarrow K^+ \pi^- \pi^-$
\\(plus charge conjugate process);
\item $B^0 \rightarrow \mu^+ \nu D^{*-} X$, 
\\with $D^{*-} \rightarrow \bar{D}^0 \pi^-, \bar{D}^0 \rightarrow K^+ \pi^-$
\\(plus charge conjugate process);
\end{enumerate}
The two channels are treated separately, with each being used to extract $a^d_{\text{sl}}$, 
before the final measurements are combined.
For clarity, the two channels are respectively denoted by $\mu D$ and $\mu D^*$ throughout this paper,
with the appropriate combinations of charges implied. Charges are only explicitly shown when required to 
describe the asymmetry measurement, or to avoid possible ambiguity.


\section{\label{sec:overview}Analysis Overview}

Experimentally, the semileptonic mixing asymmetry is expressed as:
\begin{eqnarray}
a^d_{\text{sl}} &=& \frac{A - A_{\text{BG}}}{F_{B^0}^{\text{osc}}}. \label{eq:main}
\end{eqnarray}
Here, $A$ is the measured raw asymmetry, defined by:
\begin{eqnarray}
A &=& \frac{N_{\mu^+D^{(*)-}} - N_{\mu^-D^{(*)+}}} {N_{\mu^+D^{(*)-}} + N_{\mu^-D^{(*)+}}} 
            \equiv \frac{N_{\text{diff}}}{N_{\text{sum}}}, \label{eq:raw}
\end{eqnarray}
where $N_{\mu^{\pm}D^{(*)\mp}}$ is the number of reconstructed $\mu^{\pm}D^{(*)\mp}$ signal candidates.
The sum is extracted by fitting the total mass distribution, 
and the difference by fitting the difference of two charge-specific mass distributions.
The term $A_{\text{BG}}$ accounts for inherent detector-related background asymmetries,
for example due to the different reconstruction efficiencies for positively and 
negatively charged kaons. 
The denominator $F_{B^0}^{\text{osc}}$ is defined as the fraction of all $\mu D^{(*)}$ 
signal events that arise from decays of $B^0$ mesons after they have oscillated.
It is required to account for $D^{(*)}$ mesons arising from direct $B^0$ decays, 
decays of $B^{\pm}$ and $B^0_s$ mesons, or direct hadronization from $c\bar{c}$ quarks.
All background asymmetries are extracted using data-driven methods, while Monte Carlo (MC) simulation 
is used to determine the fraction of $B^0$ mesons that have undergone mixing prior to decay.

This measurement assumes that the initial production of $B^0$--$\bar{B}^0$ is symmetric, 
and that there is no asymmetry in the decays of unmixed $B^0$ or $\bar{B}^0$ mesons (that would 
imply CPT violation), and no direct CP asymmetry in the semileptonic decays to 
charm states, or the decay of these charm states to the indicated products.  
With these assumptions, any observed semileptonic asymmetry 
would have to arise due to the mixing process.

The $B^0$ meson has a mixing frequency $\Delta M_d = 0.507 \pm 0.004$~ps$^{-1}$, 
of comparable scale to the lifetime $\tau(B^0) = 1.518 \pm 0.007$~ps~\cite{pdg}. 
Hence the fraction of oscillated $B^0$ mesons is a strong function of the measured decay time. 
The proper decay length $ct$ for a particle is given by:
\begin{eqnarray}
ct &=& \frac{L}{\beta\gamma} = L \cdot \frac{cM}{p} = L_{xy} \cdot \frac{cM}{p_T} ,
\end{eqnarray}
where $\gamma$ and $\beta$ are the usual relativistic kinematic quantities; 
$p$, $M$ and $L$ are, respectively, the particle momentum, mass and decay length in the detector reference frame. 
The best precision is obtained by using the transverse quantities $L_{xy}$ and $p_T$, 
due to finer instrumentation for tracking in this plane. The transverse decay length $L_{xy}$
is the projection of the vector pointing from the production to the decay vertex of the $B$ meson
onto the $B$ meson transverse momentum direction. It can be negative due to the limited
spatial resolution of the detector.

For semileptonic decays, the missing energy due to the undetected neutrino results in
the measured transverse momentum being underestimated with respect to the actual value. 
Hence the measured variable is actually the visible proper decay length (VPDL):
\begin{eqnarray}
\text{VPDL}(B) &=& L_{xy}(B) \cdot \frac{cM(B)}{p_T(\mu D)}. 
\end{eqnarray}
The dilution $F_{B^0}^{\text{osc}}$ is a very strong function of this variable, 
increasing monotonically with VPDL. To exploit this behavior, the measurements 
of all asymmetries are performed separately in bins of VPDL($B^0$). 
These measurements are then combined for each channel to obtain the final measurement.
The selected VPDL($B^0$) bins are defined by the edges \{$-0.10$, 0.00, 0.02, 0.05, 0.10, 0.20, 0.60\}~cm.
The $\mu D^{(*)}$ signal contributions outside of this range are found to be negligible.
The first two bins in VPDL have negligible contributions from oscillated $B^0$ mesons, and are 
not included in the final $a^d_{\text{sl}}$ measurement. They represent a control region in which the 
measured raw asymmetry should be dominated by the background contribution, i.e., $A - A_{\text{BG}} \approx 0$. 

There can be significant ($\sim$1\%) asymmetries due to detector effects. 
In particular, the material and detector elements that a particle traverses are 
different for positively and negatively charged particles, as a result of the specific 
orientation of magnetic fields in the central tracking and muon detectors. 
In this analysis, such effects are removed to first-order by reweighting all events, 
such that the total weight of events collected in each of the four (solenoid, toroid) 
magnet polarity configurations is the same (see Section~\ref{sec:det}). 
Remaining asymmetries are of order 0.1\%, and are corrected using data-driven methods.

To avoid possible experimental bias, the central values of the raw asymmetries were hidden 
until all analysis methods were finalized. Initially this was achieved by randomly assigning 
all candidate charges; later, to allow the background asymmetries to be examined, the true charges were used,
but unknown offsets were added to the raw charge asymmetries.

The D0 detector is briefly described in Section~\ref{sec:det}, highlighting those features most relevant for this measurement. 
The event selection and raw asymmetry extraction are described in Sections~\ref{sec:esel}--\ref{sec:raw}.
The determination of background asymmetries is described in Section~\ref{sec:abkg}, while Section~\ref{sec:adil}
covers the extraction of the oscillated $B^0$ fraction. The results and conclusions are presented in 
Sections~\ref{sec:results}--\ref{sec:conc}.


\section{\label{sec:det}The D0 Detector}

The D0 detector has been described in detail elsewhere~\cite{d0det}.
The most important detector components for this measurement are  
the central tracking system, the muon detectors, and the magnets.

The central tracking system comprises a silicon microstrip tracker (SMT) 
and a central fiber tracker (CFT), both located within a 2$\ts$T superconducting solenoidal magnet. 
The SMT has $\approx$$800,000$ individual strips, with typical pitch of $50-80$ $\mu$m, 
and a design optimized for tracking and vertexing capability at pseudorapidities of $|\eta|<2.5$,
where $\eta = -\text{ln}[\text{tan}(\theta/2)]$ and $\theta$ is the polar angle with respect to the beam axis.
The system has a six-barrel longitudinal structure, each with a set 
of four layers arranged axially around the beam pipe, and interspersed 
with 16 radial disks. 
In the spring of 2006, a ``Layer 0'' barrel detector with 12288 additional strips 
was installed~\cite{layer0}, and two radial disks were removed. 
This upgrade defines the chronological boundary between the two running periods, denoted Run IIa and Run IIb.
The sensors of Layer 0 
are located at a radius of 17~mm from the colliding beams.
The CFT has eight thin coaxial barrels, each 
supporting two doublets of overlapping scintillating fibers of 0.835~mm 
diameter, one doublet being parallel to the collision axis, and the 
other alternating by $\pm 3^{\circ}$ relative to the axis. Light signals 
are transferred via clear fibers to solid-state photon counters 
that have $\approx$$80$\% quantum efficiency.

A muon system resides beyond the calorimeter, and consists of a 
layer of tracking detectors and scintillation trigger counters 
before a 1.8~T toroidal magnet, followed by two similar layers after
the toroid. Tracking at $|\eta|<1$ relies on 10~cm wide drift
tubes, while 1~cm mini-drift tubes are used at
$1<|\eta|<2$.

The polarities of both the solenoidal and toroidal magnets were regularly reversed
during data acquisition, approximately every two weeks, resulting in almost
equal beam exposure in each of the four polarity configurations. 
This feature of the D0 detector is crucial in reducing detector-related asymmetries, 
for example due to the different trajectories of positive and negative muons as they
traverse the magnetic fields in the detector. 


\section{\label{sec:esel}Event Selection}

This analysis uses data collected by the D0 detector from 2002--2011, corresponding to approximately 
10.4~fb$^{-1}$ of integrated luminosity, and representing the full Tevatron Run II sample of 
$p\bar{p}$ collisions at center-of-mass energy $\sqrt{s} = 1.96$~TeV. 
Signal candidates are collected using single and dimuon triggers,
which may also impose additional criteria. 
To avoid lifetime-dependent trigger efficiencies, which are difficult to model in simulation, 
events that exclusively satisfy muon triggers with track impact-parameter requirements are removed. 

For both channels, events are considered for selection if they contain a muon candidate with reconstructed 
track segments both inside and outside the toroid magnet. 
The muon candidate must be matched to a track in the central tracking system, with at least 
three hits in both the SMT and CFT. 
In addition, it must have transverse momentum $p_T > 2$~GeV/$c$, and total momentum 
$p > 3$~GeV/$c$.

For events fulfilling these requirements, $D^{(*)\mp}$ candidates are constructed by combining 
three other tracks associated with the same initial $p\bar{p}$ interaction.  
Each track must satisfy $p_T > 0.7$~GeV/$c$, and have at least two hits in both the 
SMT and CFT.
The tracks must have a summed charge of magnitude $|q|=1$, with opposite sign to the muon charge.
Each of the tracks comprising the like-charge pair is assigned the 
charged pion mass~\cite{pdg}. The third track, which has the same charge as the muon, 
is assigned the charged kaon mass~\cite{pdg}.
 
\subsection{\label{sec:esel_mark}$\bm{\mu D}$ Channel}

For the $D^- \to K^+\pi^-\pi^-$ decay (and charge conjugate process), the three hadron tracks must be consistent 
with originating at a single common vertex, with a vertex fit to the three tracks
satisfying $\chi^2(\text{vertex}) < 16$. 
These tracks are combined to construct a $D^-$ candidate.
The resulting $D^-$ trajectory must be consistent with forming a common vertex 
with the muon to reconstruct a $B^0$ candidate.
The cosine of the angle $\theta^{D}_T$ between the momentum and trajectory vectors 
of the $D^-$ meson in the transverse plane must satisfy cos$(\theta^{D}_T) > 0.0$; 
i.e., the two vectors must point to the same hemisphere.
The invariant masses must satisfy $1.6 < M(D^-) < 2.1$~GeV/$c^2$ and $2.0 < M(B^{0}) < 5.5$~GeV/$c^2$. 

At this preselection stage, a total of $\sim$$830$ million candidates remain, 
dominated by random three-track combinations incorrectly associated with a real muon.
A fit to the $M(K\pi\pi)$ distribution yields $1\ts629\ts000 \pm 29\ts000$ $\mu D$ combinations.
To increase the signal fraction of the sample, a log likelihood ratio (LLR) method is utilized~\cite{llr1}
to construct a single discriminating parameter from the combination of thirteen individual variables:
the $D^-$ transverse decay length $L_{xy}(D^-)$, and its significance $L_{xy}(D^-) / \sigma[L_{xy}(D^-)]$;
the track isolation $I$ of the kaon, the leading pion, and the trailing pion;
the transverse momentum of the kaon, the leading pion, and the trailing pion;
the invariant mass of the reconstructed $B^0$ candidate, $M(\mu D)$;
the $\chi^2$ of the vertex fit for both the $K\pi\pi$ and $\mu D$ vertices;
and the two-dimensional angular separation $\Delta R$ of the kaon and trailing pion, 
and of the two pions.
The two-dimensional angular separation of two tracks is defined as 
$\Delta R = \sqrt{\Delta\phi^2 + \Delta\eta^2}$, where $\eta$ is the pseudorapidity 
and $\phi$ is the azimuthal angle of each track.
The track isolation $I$ is the momentum of a particle divided by the sum of momenta 
of all tracks contained in a cone of size $\Delta R=0.5$ around the particle. 
Tracks corresponding to the other three final state particles for this candidate 
are excluded from the sum.

The signal distributions required to construct the LLR discriminant are obtained from 
MC simulated events, in which the signal channel is required at generation,
and the reconstructed tracks are required to match the correct particles at the generator level.
For all MC studies described in this article, events are generated using {\sc pythia} version 6.409~\cite{pythia}, 
interfaced with {\sc evtgen}~\cite{evtgen} to model the decays of particles containing $b$ and $c$ quarks. 
Generated events are processed by a {\sc geant} based detector simulation, 
and overlaid with data from randomly collected bunch crossings to simulate pile-up 
from multiple interactions.
The MC samples are then reconstructed using the same software as used for data.
The corresponding background distributions are obtained from sideband events in real data,
defined by [$1.660 < M(K\pi\pi) < 1.760, 1.964 < M(K\pi\pi) < 2.064$]~GeV/$c^2$, with each 
sideband scaled to give equal weight to the final distributions.

Candidates enter the final data sample if the LLR discriminant exceeds 
a value $L_{\text{min}}$, chosen to maximise the signal significance in data, 
$N_S/\sqrt{N_S+N_B}$, where $N_S$ and $N_B$ are the number of $\mu D$ signal 
and background events, respectively.
This figure of merit is found to correspond to the minimum uncertainty 
on the measured raw asymmetry. The optimal requirement is determined separately in each VPDL bin,
and the value of $L_{\text{min}}$ decreases for longer lifetimes, where the background from 
random track combinations is significantly reduced.

\begin{figure}[t]
        \centering
        \includegraphics[width=\columnwidth]{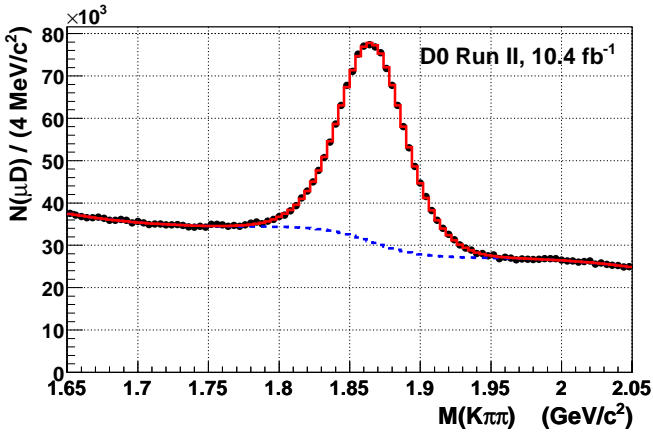}
        \caption[]{Distribution of the invariant mass $M(K\pi\pi)$ after all selections have been applied, 
for the $\mu D$ channel. The events have been weighted using the method described in Section~\ref{sec:esel_weights}. 
The histogram shows the fit model used to extract the yield, with the background component drawn 
separately as a dashed line (see Section~\ref{sec:raw} for fit models).}
\label{fig:unbinnedmass_dplus}
\end{figure}

After applying all selection requirements, the total  $\mu D$ signal yield is $\sim$$740\ts000$, 
with an overall efficiency of approximately $44$\% with respect to the preselection sample. 
The signal efficiency in VPDL bins 3--6, used to extract $a^d_{\text{sl}}$, ranges from 53\% to 72\%.
The $M(K\pi\pi)$ invariant mass distribution over the full VPDL range is shown in 
Fig.~\ref{fig:unbinnedmass_dplus}.

\subsection{\label{sec:esel_ant} $\bm{\mu D^{*}}$ Channel}

For the $D^{*-} \to \bar{D}^0\pi^-$, $\bar{D}^0 \to K^+\pi^-$ decay (and charge conjugate process), $D^{0}$ candidates are 
reconstructed by combining a pair of oppositely charged tracks passing the criteria described above.
The two tracks must form a common secondary vertex, 
and are used to reconstruct the parent $\bar{D}^0$ candidate, 
which must satisfy $p_T(\bar{D}^0) > 0.7$~GeV/$c$ and $|\eta(\bar{D}^0)| < 2.0$,
The invariant mass must lie in the range $1.7 < M(K\pi) < 2.0$~GeV/$c^2$. 

Next, an additional track is combined with the $\bar{D}^0$ candidate, 
which must have the opposite charge to the muon, and be consistent with forming a common vertex with the $\bar{D}^0$. 
This track is allocated the mass of the charged pion, and is here denoted $\pi_{D^*}$.
The difference in the invariant masses of the $D^{*-}$ and $\bar{D}^0$ candidates must satisfy 
$0.120 < [\Delta M \equiv M(K\pi\pi_{D^*}) - M(K\pi)] < 0.200$~GeV/$c^2$.
In addition, the displacement of the $\bar{D}^0 \to K^{+}\pi^{-}$ decay vertex with respect to the 
$D^{*-} \to \bar{D}^0 \pi^{-}$ decay vertex must correspond to a significance of at least $3\sigma$, i.e.,
\begin{eqnarray}
S \equiv \sqrt{(\epsilon_T/\sigma_T)^2 + (\epsilon_L/\sigma_L)^2} > 3,
\end{eqnarray}
where $\epsilon_{T(L)}$ and $\sigma_{T(L)}$ represent the distance and corresponding 
uncertainty of the transverse (longitudinal) displacement between the two vertices.

The $D^{*-}$ candidate is then combined with the muon, to form a $B^0$ candidate. 
The muon, $\bar{D}^0$, and $\pi_{D^*}$ trajectories must be consistent with arising from a common vertex, 
and the invariant mass of the $B^0$ must satisfy $2.0 < M(\mu D^{*}) < 5.5$~GeV/$c^2$. 

The final event selection requirement utilises a boosted decision tree (BDT) to further 
suppress backgrounds~\cite{tmva}. A total of 22 variables are selected as inputs:
\begin{itemize}
\item transverse momentum $p_T(K)$, $p_T(\pi)$, $p_T(\pi_{D^*})$, $p_T(\bar{D}^0)$;
\item isolation $I(K)$, $I(\pi)$, $I(\pi_{D^*})$, $I(D^{*})$, $I(B^0)$;
\item angular separation $\Delta R(K,\pi)$, $\Delta R(K,\pi_{D^*})$, $\Delta R(\pi,\pi_{D^{*}})$, $\Delta R(\bar{D}^0,\mu)$;
\item transverse decay length $L_{xy}(\bar{D}^0)$, error $\sigma[L_{xy}(\bar{D}^0)]$, 
    and significance $L_{xy}(\bar{D}^0)/\sigma[L_{xy}(\bar{D}^0)]$; 
\item cosine of the angle, in the transverse plane, between the $\bar{D}^0$ momentum vector 
  and the position vector of the $\bar{D}^0$ decay vertex with respect to 
  (a) the primary $p\bar{p}$ interaction vertex, and (b) the $B^0$ decay vertex;
\item cosine of the angle, in the transverse plane, between the $D^{*}$ momentum vector 
  and the position vector of the $D^{*}$ decay vertex with respect to the primary vertex;
\item decay vertex fit quality $\chi^2(B^0)$; and
\item invariant mass $M(K\pi)$ and $M(\mu D^{*})$.
\end{itemize}

The signal distributions are taken from MC simulation, in which the signal channel is generated 
exclusively by forcing the required decays in {\sc evtgen}, and the reconstructed tracks are required 
to match the correct particles at the generator level.
The background distributions are taken from real data, in which the kaon and two pions 
all have the same charge, and the muon has the opposite charge.
The choice of BDT cut used to define the final data sample is made separately for Run IIa and 
Run IIb samples, and for each VPDL($B^0$) region, to optimise the signal significance in each case. 

\begin{figure}[t]
        \centering
        \includegraphics[width=\columnwidth]{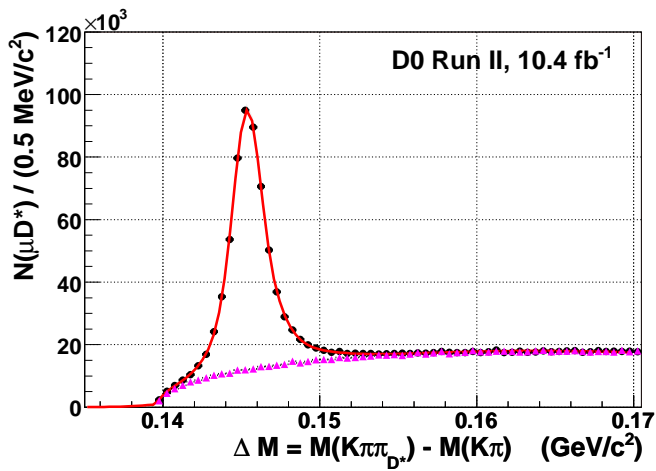}
        \caption[]{Distribution of the invariant mass difference $\Delta M \equiv [M(K\pi\pi_{D^{*}}) - M(K\pi)]$, 
for the $\mu D^{*}$ channel. The events have been weighted using the method described in Section~\ref{sec:esel_weights}. 
The solid line shows the fit model used to extract the yield. The triangular data points show the corresponding distribution for 
$\mu D^*$ candidates in which the three hadrons have the same charge, scaled to give the same yield as the 
signal sample, in the sideband region $0.155 < \Delta M < 0.170$ GeV/$c^2$ (see Section~\ref{sec:raw} for fit models).}
\label{fig:unbinnedmass_dstar}
\end{figure}

After application of all selection criteria, the sample contains $\sim$$545\ts000$ $\mu D^{*}$ signal candidates.
The $\Delta M$ distribution for the full VPDL range is shown in Fig.~\ref{fig:unbinnedmass_dstar}.

\section{\label{sec:esel_weights}Event Weights}

In any given configuration of the solenoidal and toroidal magnet polarities, there can be detector-related 
asymmetries. These originate from differing detection efficiencies for positively and negatively charged particles, 
in turn caused by their different trajectories as they bend through the magnetic fields in the detector.
The regular reversal of both magnet polarities suppresses such effects. 
To ensure maximal cancellation of these instrumental asymmetries, an additional event-by-event weighting is applied
such that the sums of weights in each (solenoid,toroid) configuration 
are the same, for a given sample.

The weights are determined after applying the final event selections, by counting the total 
number of events in each of the four solenoid and toroid magnet polarity configurations. 
The weight for an event collected in a polarity configuration $i=\{1,2,3,4\}$ is defined as $N_{\text{min}} / N_{i}$
where $N_{i}$ is the number of events in this polarity configuration, and $N_{\text{min}}$ is the smallest of the four 
yields $N_{1,2,3,4}$.
This procedure is performed separately for each channel, and for each VPDL($B^0$) bin.
Event weights are typically in the range 0.90--1.00, with very little variation between VPDL bins. 
For the unbinned samples, the total signal yields after event weighting are 
$N(\mu D) = 721\ts519 \pm 3537$, and
$N(\mu D^{*}) = 519\ts066 \pm 3446$, as shown in 
Figs.~\ref{fig:unbinnedmass_dplus} and~\ref{fig:unbinnedmass_dstar}.


\section{\label{sec:raw}Extracting the Raw Asymmetry}

The raw asymmetry is extracted by fitting the invariant mass distributions $M(K\pi\pi)$
(or $\Delta M$) for the $D^{(*)}$ candidates. 
The sum distribution $H_{\text{sum}}$ is constructed by weighting all $\mu D^{(*)}$ 
candidates according to the magnet polarity weight.
A difference distribution $H_{\text{diff}}$ is constructed by taking the difference betwen the $\mu^+ D^{(*)-}$
and $\mu^- D^{(*)+}$ distributions.
The sum and difference distributions are modeled by, respectively, the functions:
\begin{eqnarray}
F_{\text{sum}}  & = & F^{\text{BG}}_{\text{sum}}  +  N_{\text{sum}} \cdot F^{\text{sig}} , \\
F_{\text{diff}} & = & F^{\text{BG}}_{\text{diff}} + A \cdot N_{\text{sum}} \cdot F^{\text{sig}},
\end{eqnarray}
where $N_{\text{sum}}$ is the total $\mu D^{(*)}$ yield, and $A$ is the 
corresponding raw charge asymmetry defined in Eq.~(\ref{eq:raw}). 
Different models are used to parametrize the backgrounds for the sum ($F^{\text{BG}}_{\text{sum}}$) 
and difference ($F^{\text{BG}}_{\text{diff}}$) histograms, 
while a single model $F^{\text{sig}}$ is used for the signal in both cases.
The yields, asymmetries, and signal and background parameters in these models are 
extracted by a simultaneous binned fit to the two distributions, 
to minimise the total $\chi^2$ with respect to the fitting functions.

\subsection{\label{sec:raw_mark}$\bm{\mu D}$ Channel}

The physical width of the $D^{-}$ meson is negligible compared to the detector resolution;
therefore, the signal parametrization is chosen based on studies of simulated data to determine the 
mass resolution for this channel. 
The signal is modeled by the sum of two Gaussian functions constrained to have the same mean, 
but with different widths and relative normalizations: 
\begin{eqnarray}
F^{\text{sig}}(D)  =  \frac{1}{\sqrt{2\pi}} &\bigg [& f_{G1}  \cdot \frac{1}{\sigma_1} \cdot e^{-(x - M_D)^2/(2\sigma_1^2)} \\
                                               &+& (1 - f_{G1}) \cdot \frac{1}{\sigma_2} \cdot e^{-(x - M_D)^2/(2\sigma_2^2)} \bigg ], \nonumber
\end{eqnarray}
where $x$ is the reconstructed invariant mass of the $D^{-}$ candidate, 
$M_D$ is the mean of the Gaussian peak, 
$\sigma_{1,2}$ are the widths of the first and second Gaussians, 
and $f_{G1}$ is the fraction of the signal in the first Gaussian peak.

The background in the sum distribution exhibits slightly different behavior 
for each VPDL bin, hence a flexible parametrization is selected to provide 
good agreement in all bins, comprising the sum of three possible components: 
a low-order polynomial function, 
a falling exponential function, 
and a hyperbolic tangent function:
\begin{eqnarray}
F^{\text{BG}}_{\text{sum}}(D)  =  \label{eq:bgsum} \\
 (N_{\text{tot}} - N_{\text{sum}}) \cdot &\{& f_{\text{tanh}}  \cdot F_{\text{tanh}}  \nonumber \\
                                    &+& ( 1 - f_{\text{tanh}}) \cdot f_{\text{poly}}  \cdot F_{\text{poly}} \nonumber \\
                                    &+& ( 1 - f_{\text{tanh}}) \cdot ( 1 - f_{\text{poly}}) \cdot F_{\text{exp}} \}. \nonumber
\end{eqnarray}
Here $f_{\text{tanh}}$ and $f_{\text{poly}}$ are free parameters between zero and one, 
controling the relative contributions of the three components.

The polynomial function includes a constant, linear, and cubic term:
\begin{eqnarray}
F_{\text{poly}} &=& C_1  + p_1 \cdot (x - x_{\text{mid}}) + p_3 \cdot (x - x_{\text{mid}})^3 ,
\end{eqnarray}
where $p_1$ and $p_3$ are free parameters of the fit, and $x_{\text{mid}}$ is the 
mid-point of the fitting range. The exponential term is:
\begin{eqnarray}
F_{\text{exp}} &=& C_2  \cdot e^{r (x - x_{\text{min}})} ,
\end{eqnarray}
where $r$ is a free parameter and $x_{\text{min}}$ is the lower-limit of the fitting range.

Finally, a hyperbolic tangent function described by
\begin{eqnarray}
F_{\text{tanh}} &=& C_3 \cdot [1 - \text{tanh}(k \cdot(x - M_D))] \label{eq:tanh}
\end{eqnarray}
is used to model the effects of partially reconstructed decays, and reflections from decays 
into other three-track combinations. Monte Carlo simulations confirm that this 
parametrization is a good model for this source of background, which includes contributions from
$D^{-}$ decays to $K^{+}\pi^{-}\pi^{-}\pi^0$, $\pi^{+}\pi^{-}\pi^{-}\pi^0$, and $K^{+}K^{-}\pi^{-}$; 
$\bar{D}^0$ decays to four charged hadrons; and decays of $D^{*-} \to \bar{D}^0\pi^{-}$, with $\bar{D}^0 \to K^{+}\pi^{-}\pi^0$,
where the $\pi^0$ is not reconstructed. 
The steepness of the threshold, denoted by $k$ in Eq.~(\ref{eq:tanh}), 
is controlled by the detector mass resolution. As such it is fixed according 
to the widths of the two Gaussian peaks:
\begin{eqnarray}
k &=& \frac{1}{\sqrt{2\sigma_{\text{mean}}^2}},
\end{eqnarray}
where $\sigma_{\text{mean}} = f_{G1}\cdot\sigma_1 + (1 - f_{G1})\cdot\sigma_2$ 
is the weighted mean of the two widths.

All three individual components are normalized to have unit area in the fitting range, 
by suitable choice of the constants $C_1$, $C_2$, and $C_3$. 
The overall normalization is set by subtracting the fitted number of signal events 
($N_{\text{sum}}$) from the total event count in the sample ($N_{\text{tot}}$). 
Using the total event count as a constraint in this way improves the fit precision. 
The free parameters are the two fractions, $f_{\text{tanh}}$ and $f_{\text{poly}}$, 
the two polynomial coefficients, $p_{1,3}$, and the argument of the exponential function, $r$. 
This empirical choice provides good agreement with the data, with relatively few 
free parameters, over a range of different background shapes.
To improve fit precision and stability, each term in $F^{\text{BG}}_{\text{sum}}$ 
is only used if it improves the fit probability. 
As a result, for VPDL bins 1--3 the exponential component is removed; 
for bin 6, the cubic term is removed.

For the difference fit, the overall normalization $N_{\text{diff}}^{\text{BG}}$ 
of the background is fixed according to the observed number of $D^+$ and $D^-$ events, 
and the signal contribution:
\begin{eqnarray}
N_{\text{diff}}^{\text{BG}} = N^+_{\text{tot}} - N^-_{\text{tot}} - A \cdot N_{\text{sum}}, \label{eq:dplus_bgdiff1}
\end{eqnarray}
where $N^{\pm}_{\text{tot}}$ is the sum of all event weights for $D^{\pm}$ candidates. 
The background model comprises the same three components as used in the fit to the sum. 
The threshold component is modeled by the same shape as in the sum fit, 
with the yield scaled by a free parameter ($a_{\text{tanh}}$) accounting for the 
possible charge asymmetry from this contribution.
The combinatorial background is modeled by the sum of exponential and polynomial terms, 
with the shape parameters common to the sum fit, and the yield constrained by 
Eq.~(\ref{eq:dplus_bgdiff1}) after the contribution of the threshold component has been accounted for:
\begin{eqnarray}
F^{\text{BG}}_{\text{diff}}(D)  = \label{eq:bgdiff} \\
       N_{\text{diff}}^{\text{BG}}  \cdot &\{& a_{\text{tanh}} \cdot f_{\text{tanh}}  \cdot F_{\text{tanh}} \nonumber  \\
                                             &+& ( 1 - a_{\text{tanh}} \cdot f_{\text{tanh}}) \cdot f_{\text{poly}}  \cdot F_{\text{poly}} \nonumber \\
                                             &+& ( 1 - a_{\text{tanh}} \cdot f_{\text{tanh}}) \cdot ( 1 - f_{\text{poly}}) \cdot F_{\text{exp}} \}. \nonumber
\end{eqnarray}
This function has only one additional free parameter, with respect to the fit 
over the sum of all candidates, namely the asymmetry on the hyperbolic tangent, $a_{\text{tanh}}$.
The corresponding asymmetry term for the polynomial component is eliminated by applying 
the constraint from Eq.~(\ref{eq:dplus_bgdiff1}). 

In total, there are twelve free parameters in the mass fit for this channel, 
six describing the signal, and six describing the background. 
The default fit is performed over the range $[1.65 < M(K\pi\pi) < 2.05]$ GeV/$c^2$, 
using 100 bins of width 4~MeV/$c^2$, with variations on both the fitting range and the 
bin width considered as sources of systematic uncertainty.

\subsection{\label{sec:raw_ant} $\bm{\mu D^{*}}$ Channel}

For this channel the invariant mass difference distribution $\Delta M = M(K\pi\pi_{D^*}) - M(K\pi)$ 
is fitted to extract the raw asymmetry.
The proximity to the pion production threshold at approximately $140$~MeV/$c^2$ leads to phase-space effects 
that tend to distort the signal and background distributions. 
To account for these effects, and based on studies of MC simulation data, the signal is modeled by a skewed triple-Gaussian function, 
i.e., three Gaussian peaks constrained to have the same mean, but with different widths and relative contributions, 
and each multiplied by a threshold shape:
\begin{eqnarray}
F^{\text{sig}}(D^*)       =  f_{G1}       &\cdot&  G_{\text{sk}}(x,\sigma_1,M,s)  \\
        + (1 - f_{G1}) \cdot f_{G2}       &\cdot&  G_{\text{sk}}(x,\sigma_2,M,s)  \nonumber \\
        + (1 - f_{G1}) \cdot (1 - f_{G2}) &\cdot&  G_{\text{sk}}(x,\sigma_3,M,s) \nonumber
\end{eqnarray}
where $G_{\text{sk}}$ is the skewed Gaussian function:
\begin{eqnarray}
G_{\text{sk}}(x,\sigma_i,M,s) = \frac{1}{\sqrt{2\pi}\sigma_i} \cdot E[-s\frac{(x-M)}{\sqrt{2}\sigma_i}] \cdot e^{-\frac{(x - M)^2}{2\sigma_i^2}}. \nonumber \\
\end{eqnarray}
Here $x$ is the reconstructed value of $\Delta M$ for this $D^{*-}$ candidate, 
$M$ is the mean of the Gaussian peak, $\sigma_{1,2,3}$ are the widths of the three Gaussians, 
and $f_{G1,2}$ describe their relative fractional contributions. 
The function $E(s \cdot y)$ is a threshold shape modeled by the complementary error function, 
taking the skew $s$ as an input parameter, and defined as: 
\begin{eqnarray}
E(s \cdot y) = \frac{2}{\sqrt{\pi}} \int_{s \cdot y}^\infty e^{-t^2} dt .
\end{eqnarray}

The background in the sum fit is modeled by the product of a linear function 
and a power law function with a threshold at the charged pion mass $M_{\pi}$~\cite{pdg},
and three free parameters $a$, $b$, and $d$: 
\begin{eqnarray}
F^{\text{BG}}_{\text{sum}}(D^*)  =  d \cdot (x - M_{\pi})^a \cdot (1 + bx) \label{eq:bgsum_ant},
\end{eqnarray}
The background in the difference distribution is modeled by the same shape as used for the sum, 
but with a different overall scale, quantified by a background asymmetry parameter $a_{\text{BG}}$:
\begin{eqnarray}
F^{\text{BG}}_{\text{diff}}(D^*)  =  d \cdot a_{\text{BG}} \cdot (x - M_{\pi})^a \cdot (1 + bx).
\end{eqnarray}

In total there are thirteen free parameters in the fit, nine for the signal, and four describing the background.
The default fit is performed over the range $[0.139 < \Delta M < 0.170]$ GeV/$c^2$, using 62 bins of 
width 0.5~MeV/$c^2$, with variations on both the fitting range and the bin width considered as 
sources of systematic uncertainty.

\subsection{\label{sec:raw_results} Results}

\begin{figure*}[t]
        \centering
	\subfigure[~$H_{\text{sum}}$ ($\mu D$ channel)]
        {\includegraphics[width=\columnwidth]{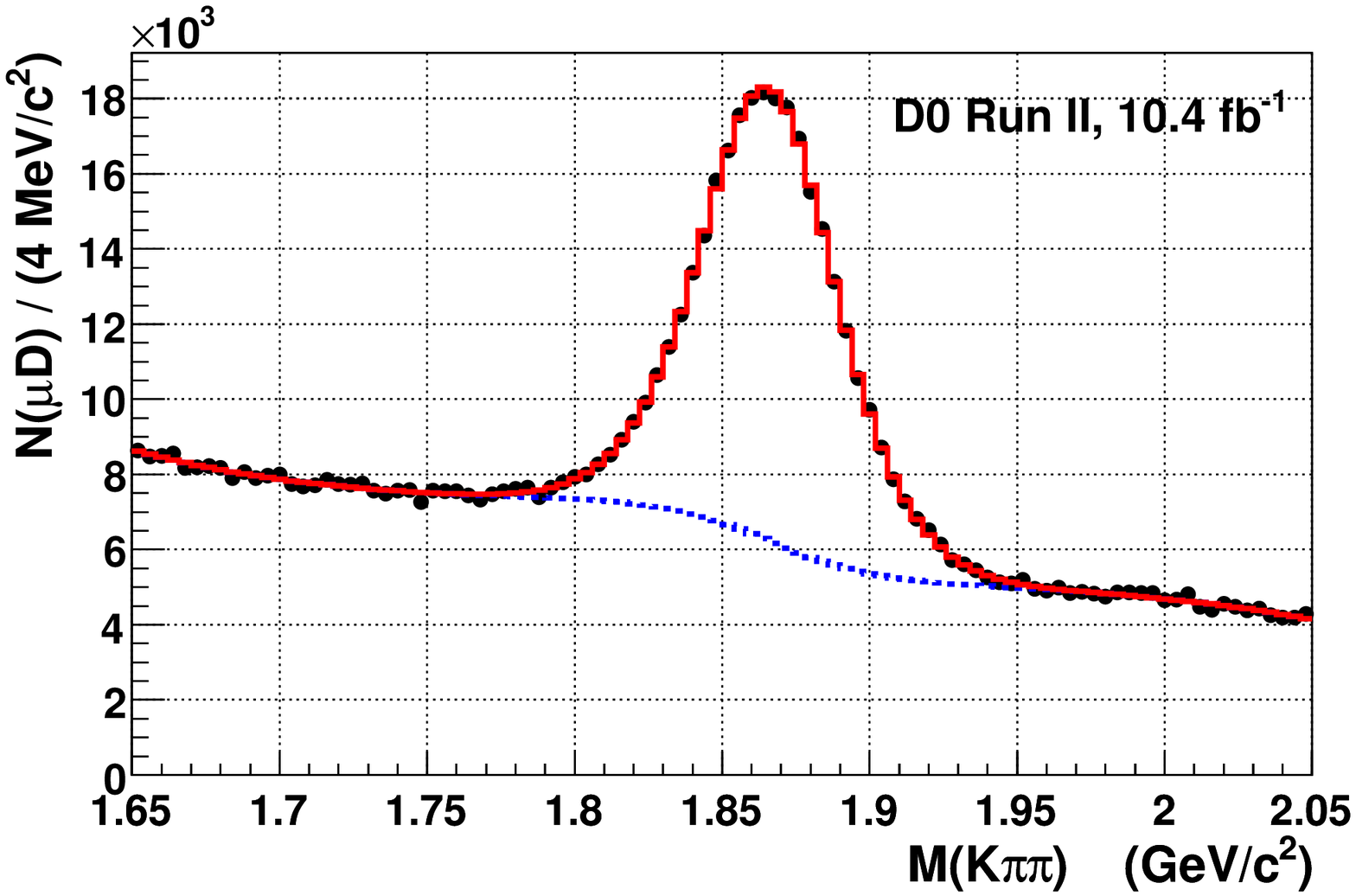}}
        \subfigure[~$H_{\text{diff}}$ ($\mu D$ channel)]
        {\includegraphics[width=\columnwidth]{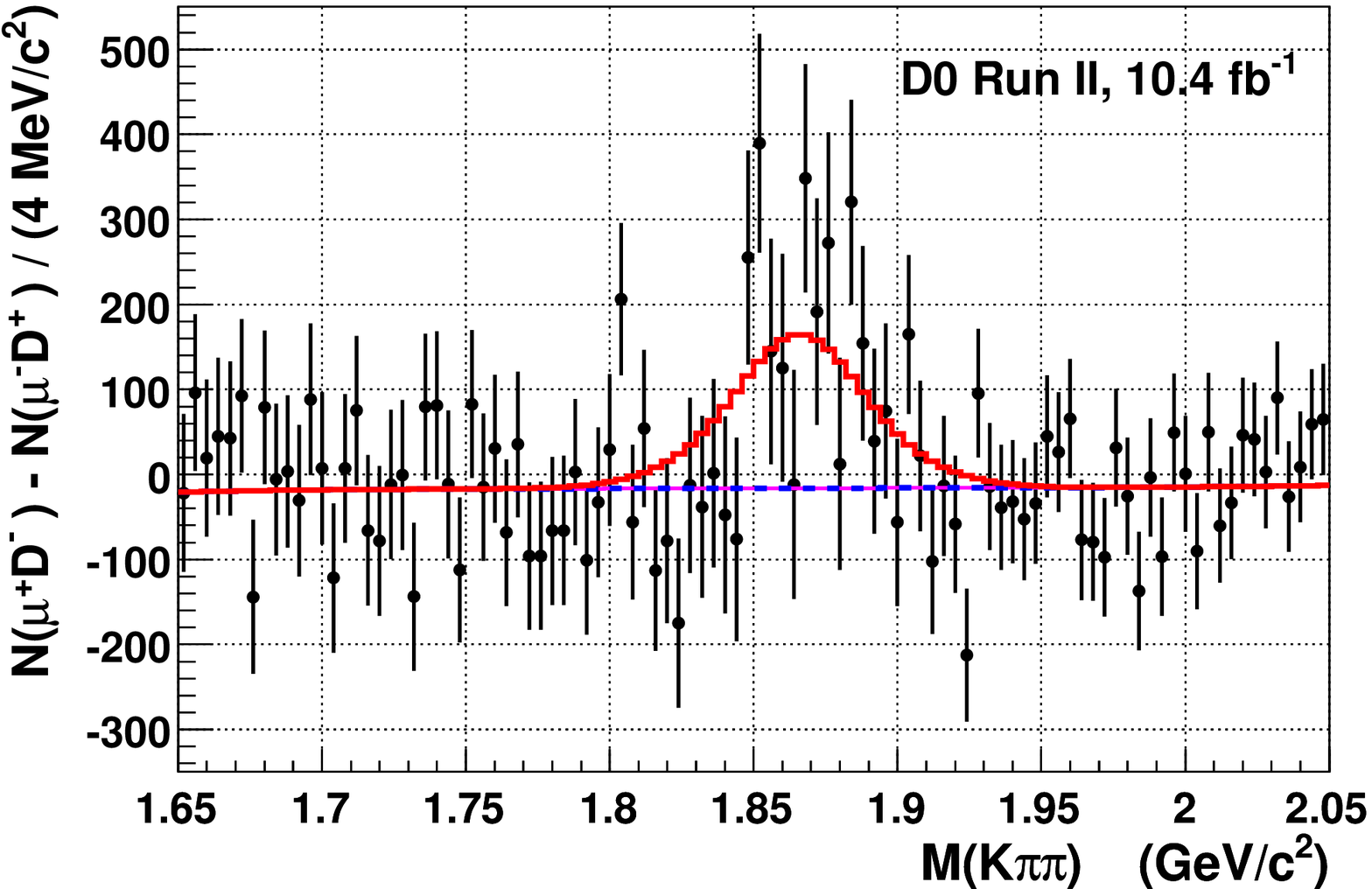}}
	\subfigure[~$H_{\text{sum}}$ ($\mu D^*$ channel)]
        {\includegraphics[width=\columnwidth]{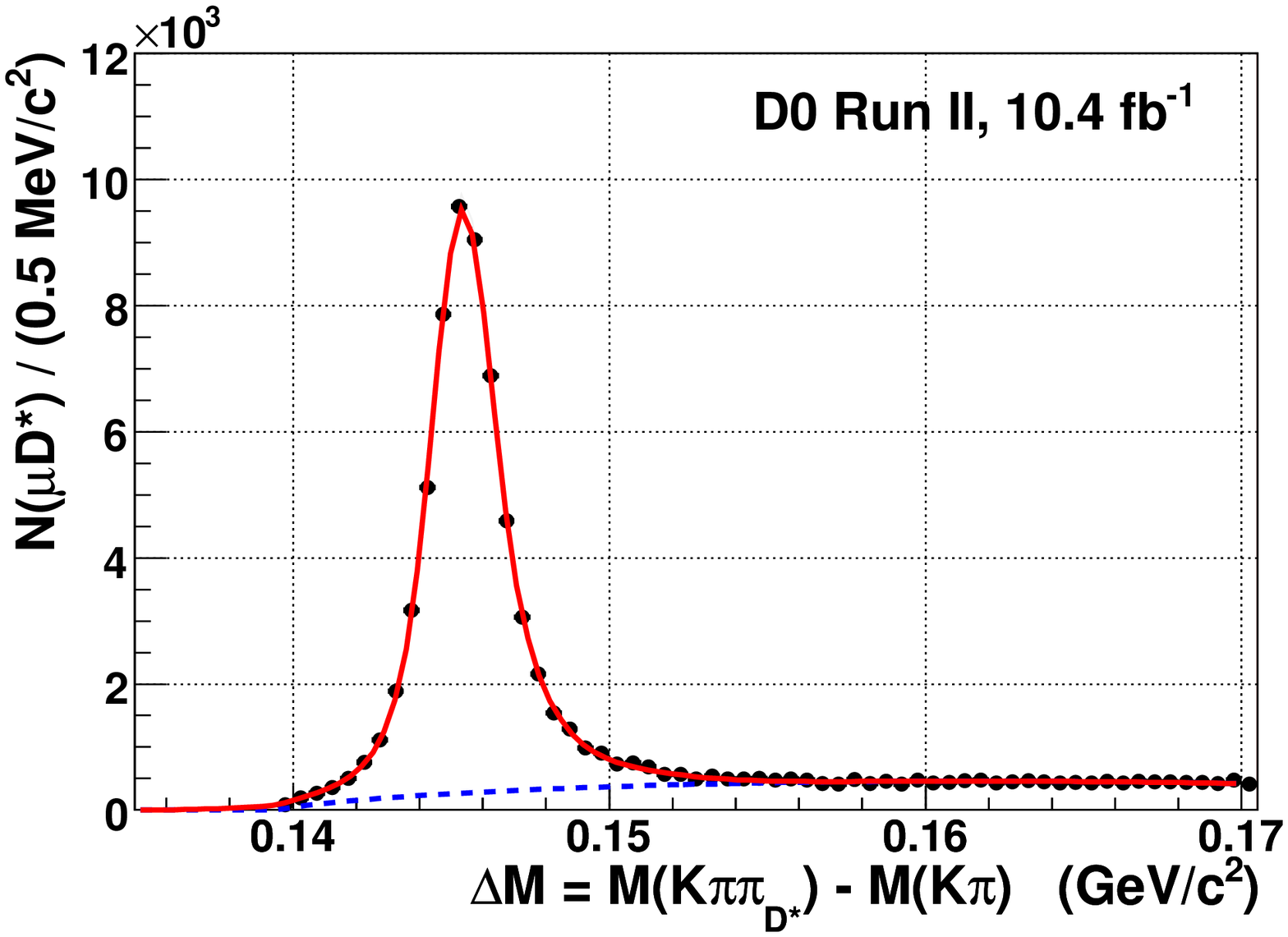}}
        \subfigure[~$H_{\text{diff}}$ ($\mu D^*$ channel)]
        {\includegraphics[width=\columnwidth]{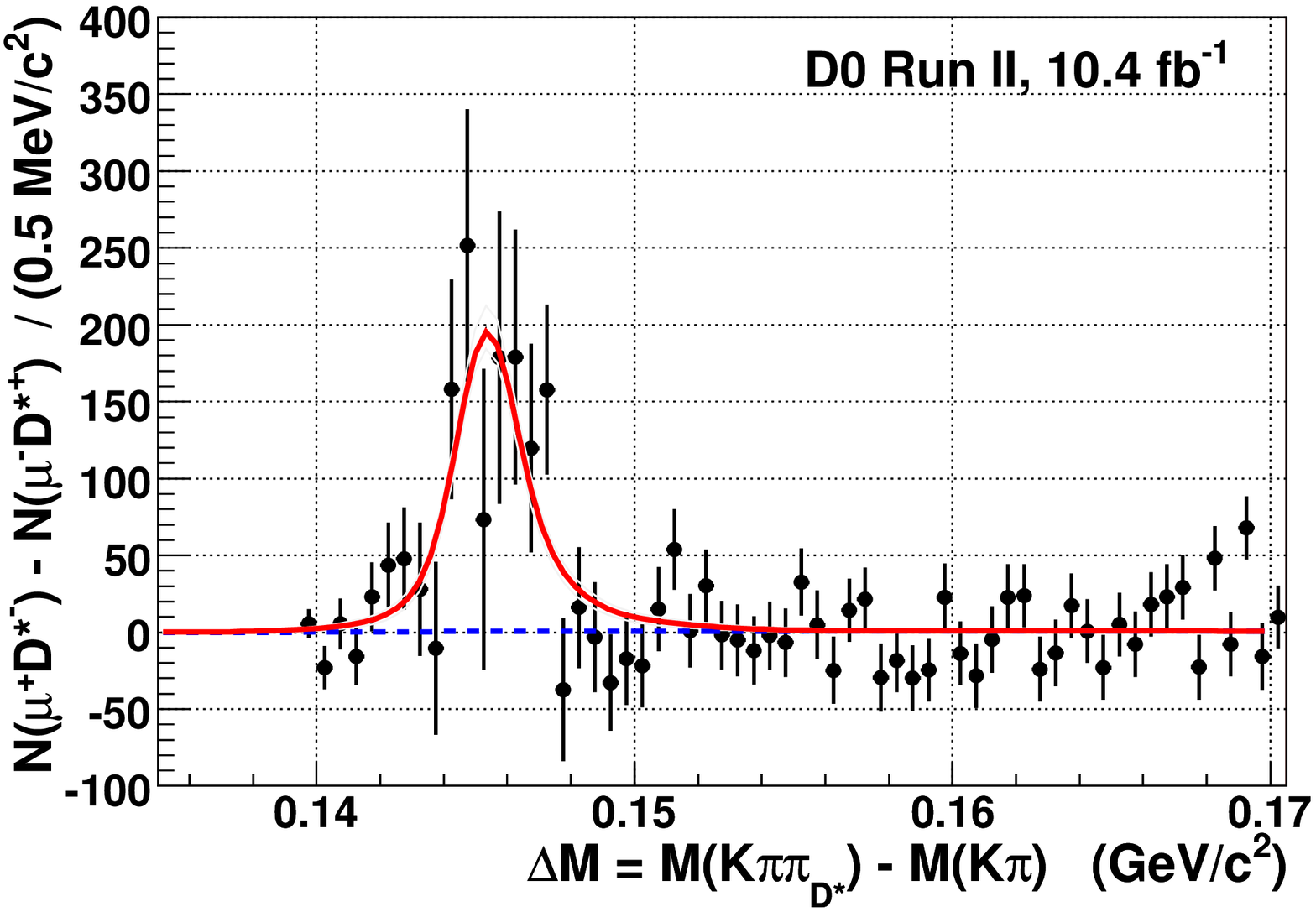}}
        \caption[]{Examples of the raw asymmetry fit for the two decays channels, 
          for the fifth VPDL($B^0$) bin corresponding to ($0.10 < \text{VPDL}(B^0) < 0.20$)~cm.
          The left plots show the sum distributions; the right plots show the difference distributions. 
          In both cases, the solid line represents the total fit function, with the background part 
          shown separately by the dashed line (see text).}
\label{fig:rawresults}
\end{figure*}

The values of all physics parameters returned by the fits, for both channels, 
and for each of the six VPDL($B^0$) bins are collected in 
Tables~\ref{tab:rawresults_mark}--\ref{tab:rawresults_ant}, with examples of the fit 
projections shown in Fig.~\ref{fig:rawresults}.
Significant positive asymmetries are observed for all VPDL bins, including those in the control region
$\text{VPDL}(B^0) < 0.02$~cm. This is expected as a consequence of the positive kaon reconstruction asymmetry,
which is described and corrected for in Section~\ref{sec:abkg}.
The two channels have similar statistical precision on the raw asymmetries, except for the first VPDL bin, 
where the sensitivity of the $\mu D$ channel is significantly reduced by the increased background from random 
three track combinations close to the primary $p\bar{p}$ interaction. The $\mu D^*$ channel is less susceptible 
to such effects, due to the intermediate resonance in the decay.

\renewcommand\arraystretch{1.1}
\begin{table*}[t]
\caption[]
{\label{tab:rawresults_mark}Results of the raw asymmetry fits for the $\mu D$ channel, 
in each of the six bins of VPDL($B^0$). The uncertainties are statistical, as returned by the fits.}
\begin{center}
\begin{tabular}{|l||rcl|rcl|rcl|rcl|rcl|rcl|}
\hline \hline
                           & \multicolumn{3}{c|}{Bin 1} &  \multicolumn{3}{c|}{Bin 2}  &  \multicolumn{3}{c|}{Bin 3}
                             & \multicolumn{3}{c|}{Bin 4} &  \multicolumn{3}{c|}{Bin 5}  &  \multicolumn{3}{c|}{Bin 6} \\
\hline
VPDL($B^0$) (cm)           & $-0.10$ & -- & $0.00$   & $0.00$ & -- & $0.02$     & $0.02$ & -- & $0.05$   
                             & $0.05$ & -- & $0.10$   & $0.10$ & -- & $0.20$     & $0.20$ & -- & $0.60$       \\ 
\hline \hline
$N(\mu D)$                 & $42\,707$ & $\pm$ & $1374$ & $155\,322$ & $\pm$ & $1011$  & $198\,874$ & $\pm$ & $1105$   
                             & $182\,921$ & $\pm$ & $1598$   & $113\,965$ & $\pm$ & $1329$     & $26\,939$ & $\pm$ & $458$    \\ 
\hline
$A$ (\%)                   & 2.70 & $\pm$ & 1.28         & 1.02 & $\pm$ & 0.35           & 1.16 & $\pm$ & 0.32   
                             & 1.50 & $\pm$ & 0.33         & 1.48 & $\pm$ & 0.41           & 1.20 & $\pm$ & 0.88       \\ 
\hline
$M(D)$ (MeV/$c^2$)         & $1866.3$ & $\pm$ & $0.6$   & $1865.8$ & $\pm$ & $0.2$     & $1865.7$ & $\pm$ & $0.2$   
                             & $1865.9$ & $\pm$ & $0.2$   & $1865.4$ & $\pm$ & $0.2$     & $1864.3$ & $\pm$ & $0.3$    \\ 
\hline
$\sigma_{G1}$ (MeV/$c^2$)  & $21.6$ & $\pm$ & $1.6$     & $19.0$ & $\pm$ & $0.9$       & $18.6$ & $\pm$ & $1.0$   
                             & $18.3$ & $\pm$ & $1.0$     & $18.0$ & $\pm$ & $1.7$       & $15.0$ & $\pm$ & $3.0$      \\ 
\hline
$\sigma_{G2}$ (MeV/$c^2$)  & $39.6$ & $\pm$ & $8.8$     & $32.5$ & $\pm$ & $1.4$       & $30.1$ & $\pm$ & $1.1$   
                             & $29.6$ & $\pm$ & $1.3$     & $28.6$ & $\pm$ & $1.8$       & $27.0$ & $\pm$ & $1.5$      \\ 
\hline
$f_{G1}$                   & $0.67$ & $\pm$ & $0.16$    & $0.417$ & $\pm$ & $0.071$    & $0.361$ & $\pm$ & $0.076$   
                             & $0.363$ & $\pm$ & $0.084$    & $0.33$ & $\pm$ & $0.13$    & $0.16$ & $\pm$ & $0.11$     \\ 
\hline
$a_{\text{tanh}}$ (\%)     & $3.75$ & $\pm$ & $2.86$    & $1.32$ & $\pm$ & $1.04$    & $-0.02$ & $\pm$ & $0.84$   
                             & $-0.04$ & $\pm$ & $0.79$    & $-0.24$ & $\pm$ & $1.00$    & $-0.99$ & $\pm$ & $3.66$     \\ 
\hline \hline
$\chi^2$/ndf               & $172$ & $/$ & $190$        & $199$ & $/$ & $190$          & $218$ & $/$ & $190$   
                             & $199$ & $/$ & $188$         & $213$ & $/$ & $188$         & $170$ & $/$ & $189$       \\ 
\hline
$\chi^2$(sum)/ndf          & $81$ & $/$ & $92$          & $105$ & $/$ & $92$           & $114$ & $/$ & $92$   
                             & $108$ & $/$ & $90$         & $123$ & $/$ & $90$           & $79$ & $/$ & $91$          \\ 
\hline
$\chi^2$(diff)/ndf         & $91$ & $/$ & $98$          & $94$ & $/$ & $98$            & $104$ & $/$ & $98$   
                             & $92$ & $/$ & $98$         & $90$ & $/$ & $98$             & $91$ & $/$ & $98$          \\ 
\hline \hline
\end{tabular}
\end{center}
\end{table*}
\renewcommand\arraystretch{1.0}

\renewcommand\arraystretch{1.1}
\begin{table*}[t]
\caption[]
{\label{tab:rawresults_ant}Results of the raw asymmetry fits for the $\mu D^*$ channel, 
in each of the six bins of VPDL($B^0$). The uncertainties are statistical, as returned by the fits.}
\begin{center}
\begin{tabular}{|l||rcl|rcl|rcl|rcl|rcl|rcl|}
\hline \hline
                           & \multicolumn{3}{c|}{Bin 1} &  \multicolumn{3}{c|}{Bin 2}  &  \multicolumn{3}{c|}{Bin 3}
                             & \multicolumn{3}{c|}{Bin 4} &  \multicolumn{3}{c|}{Bin 5}  &  \multicolumn{3}{c|}{Bin 6}  \\
\hline
VPDL($B^0$) (cm)           & $-0.10$ & -- & $0.00$   & $0.00$ & -- & $0.02$     & $0.02$ & -- & $0.05$   
                             & $0.05$ & -- & $0.10$   & $0.10$ & -- & $0.20$     & $0.20$ & -- & $0.60$        \\ 
\hline \hline
$N(\mu D^*)$               & 59\,823 &$\pm$& 2398         & 151\,585 &$\pm$& 1677         & 132\,227 &$\pm$& 1092 
                                    & 104\,463 &$\pm$& 921        & 58\,409 &$\pm$& 651           & 12\,029 &$\pm$& 233              \\
\hline 
$A$ (\%)                   & 1.82 &$\pm$& 0.67          & 1.10 &$\pm$& 0.30            & 0.94 &$\pm$& 0.30 
                                    & 1.38 &$\pm$& 0.33          & 2.11 &$\pm$& 0.44            & 0.55 &$\pm$& 0.99             \\

\hline 
$M$ (MeV/$c^{2}$)          & 145.08 &$\pm$& 0.06        & 145.07 &$\pm$& 0.02         & 145.03 &$\pm$& 0.02 
                                    & 144.99 &$\pm$& 0.02       & 144.97 &$\pm$& 0.03         & 145.00 &$\pm$& 0.06            \\
\hline 
$\sigma_{G1}$ (MeV/c$^{2}$) & 0.76 &$\pm$& 0.11         & 1.59 &$\pm$& 0.12         & 1.72 &$\pm$& 0.09 
                                    & 1.75 &$\pm$& 0.12          & 1.67 &$\pm$& 0.15          & 1.39 &$\pm$& 0.23            \\
\hline 
$\sigma_{G2}$ (MeV/c$^{2}$) & 1.45 &$\pm$& 0.30         & 0.84 &$\pm$& 0.03         & 0.86 &$\pm$& 0.02 
                                    & 0.90 &$\pm$& 0.03          & 0.90 &$\pm$& 0.04          & 0.79 &$\pm$& 0.11            \\
\hline 
$\sigma_{G3}$ (MeV/c$^{2}$) & 3.78 &$\pm$& 0.76         & 3.45 &$\pm$& 0.30         & 3.89 &$\pm$& 0.31 
                                    & 3.87 &$\pm$& 0.32          & 3.74 &$\pm$& 0.30          & 2.92 &$\pm$& 0.36            \\
\hline 
$f_{G1}$                    & 0.32 &$\pm$& 0.14         & 0.39 &$\pm$& 0.03         & 0.41 &$\pm$& 0.02 
                                    & 0.38 &$\pm$& 0.02          & 0.38 &$\pm$& 0.04          & 0.44 &$\pm$& 0.10            \\
\hline 
$f_{G2}$                    & 0.64 &$\pm$& 0.05         & 0.69 &$\pm$& 0.05         & 0.74 &$\pm$& 0.04 
                                    & 0.74 &$\pm$& 0.04          & 0.69 &$\pm$& 0.06          & 0.52 &$\pm$& 0.17            \\
\hline
$s$                         & 0.368 &$\pm$& 0.083       & 0.470 &$\pm$& 0.032       & 0.510 &$\pm$& 0.024 
                                    & 0.541 &$\pm$& 0.025       & 0.574 &$\pm$& 0.037         & 0.508 &$\pm$& 0.079        \\
\hline 
$a_{\text{BG}}$ (\%)      & 1.25 &$\pm$& 0.16         & 1.17 &$\pm$& 0.22         & $-$0.10 &$\pm$& 0.39 
                                    & $-$0.19 &$\pm$& 0.53       & 0.18 &$\pm$& 0.77          & $-$0.94 &$\pm$& 1.41        \\
\hline\hline 
$\chi^{2}$/ndf              & 133 &/& 113               & 138 &/& 113               & 203 &/& 107                                                                                                      & 153 &/& 107                & 165 &/& 107                & 150 &/& 107                    \\ 
\hline 
$\chi^{2}$(sum)/ndf         & 80 &/& 53                 & 84 &/& 53                 & 159 &/& 53                                                                                                       & 94 &/& 53                  & 94 &/& 53                  & 77 &/& 53                      \\ 
\hline 
$\chi^{2}$(diff)/ndf        & 52 &/& 60                 & 54 &/& 60                 & 45 &/& 60                                                                                                        & 59 &/& 60                  & 71 &/& 60                  & 73 &/& 60                      \\ 
\hline \hline
 \end{tabular} 
 \end{center} 
 \end{table*}
\renewcommand\arraystretch{1.0}

The raw asymmetry measurement method is validated by the use of ensemble tests, 
in which the fits to data are repeated several thousand times with the $\mu^{\pm}D^{(*)\mp}$ 
charges randomized independently for each fit.
Different input asymmetries are simulated, ranging from $-5$\% to $+5$\%, and the distribution of 
asymmetries extracted from the fits are examined.
For all cases, the distributions are well-modeled by Gaussian peaks, with a central value consistent 
with the input asymmetry, and a width consistent with the corresponding uncertainty reported in 
Tables~\ref{tab:rawresults_mark}--\ref{tab:rawresults_ant}.
A similar approach is used to confirm that the optimal precision is obtained by maximizing the signal significance.

\subsection{\label{sec:raw_syst} Systematic Uncertainties}

Several variations of the fits are performed to extract the raw asymmetry from the data, 
in order to examine the resulting spread of measured values, and assign an appropriate 
systematic uncertainty. These variations are considered for all cases where there is a 
reasonable alternative to the choices made for the nominal fit, namely:
\begin{itemize}
\item the lower and upper limits of the fitting region are varied to a number of different 
options within a 50~MeV/$c^2$ (10~MeV/$c^2$) range of the nominal choice for the $\mu D^{(*)}$ case;
\item the bin width is varied, with alternative widths of 2--20~MeV/$c^2$ for 
the $\mu D$ case, and 0.5--2.0~MeV/$c^2$ for the $\mu D^*$ case;
\item for the $\mu D$ channel, the function $F^{\text{BG}}_{\text{sum}}$ 
used to model the background of the sum distribution is changed to three alternative models in 
addition to the nominal choice; in one alternative, the polynomial is fixed to a linear function; 
in another, the polynomial is set to a quadratic function; in the third alternative, 
the mean of the hyperbolic tangent is allowed to be a free parameter, 
rather than being constrained to the mean $D^{-}$ mass;
\item for the $\mu D^*$ channel, the mass window used to select 
$D^0 \to K\pi$ candidates is varied, from the nominal requirement of 
$1.7 < M(K\pi) < 2.0$~GeV/$c^2$, to twelve alternative ranges, 
giving different background fractions and shapes in the final distribution;
\item the function $F^{\text{sig}}$, used to model the signal shape, 
is changed to an alternative choice; 
for the $\mu D$ case, a single Gaussian function is used;
for the $\mu D^*$ case, a skewed double-Gaussian is used;
\item the function $F^{\text{BG}}_{\text{diff}}$, used to model the background of the 
difference distribution, is changed to an alternative choice; 
for the $\mu D$ case, a linear function is used;
for the $\mu D^*$ case, a second or fourth order polynomial is used;
\item the event weights are allocated using an alternative method, 
based on the fitted number of $\mu D^{(*)}$ signal events in each polarity configuration,
rather than the total number of candidates.
\end{itemize}
To properly assess the combined effect of all these adjustments to the fit, 
including correlations, all possible combinations of the above fit variations are tested,
and the systematic uncertainty is allocated as the standard deviation of the full set of alternative measurements.
Table~\ref{tab:syst} shows the final systematic uncertainties allocated for each VPDL bin, for both channels. 
The combined systematic uncertainty is significantly smaller than the statistical uncertainty in all cases.

\renewcommand\arraystretch{1.1}
\begin{table*}[t]
\caption[]
{\label{tab:syst}Systematic uncertainties on the raw asymmetry measurement for both channels, 
extracted from an ensemble of fits with variations on each quantity tested. 
Also shown are the corresponding statistical uncertainties, for comparison.}
\begin{center}
\begin{tabular}{|l||c|c|c|c|c|c|}
\hline \hline
\multirow{2}{*}{Source}      
       & Bin 1               & Bin 2               & Bin 3               & Bin 4               & Bin 5                & Bin 6               \\
       & $-0.10$ -- $0.00$~cm & $0.00$ -- $0.02$~cm  & $0.02$ -- $0.05$~cm  & $ 0.05$ -- $0.10$~cm & $0.10$ -- $0.20$~cm   & $0.20$ -- $0.60$~cm  \\ 
\hline \hline
\multicolumn{7}{|c|}{$\mu D$ channel}  \\
\hline
Bin width     
       & $0.09\%$       & $0.01\%$        & $0.01\%$        & $0.01\%$         & $0.00\%$          & $0.05\%$      \\ 

Fit limits     
       & $0.17\%$       & $0.06\%$        & $0.08\%$        & $0.05\%$         & $0.03\%$          & $0.12\%$      \\ 

Magnet weighting    
       & $0.02\%$       & $0.00\%$        & $0.00\%$        & $0.00\%$         & $0.00\%$          & $0.01\%$      \\ 

Signal model     
       & $0.03\%$       & $0.03\%$        & $0.01\%$        & $0.04\%$         & $0.01\%$          & $0.01\%$      \\ 

Background model (sum)     
       & $0.03\%$       & $0.00\%$        & $0.01\%$        & $0.01\%$         & $0.01\%$          & $0.00\%$      \\ 

Background model (diff)     
       & $0.01\%$       & $0.00\%$        & $0.01\%$        & $0.00\%$         & $0.01\%$          & $0.02\%$      \\ 
\hline 
Combined systematic     
       & $\pm 0.19\%$       & $\pm 0.07\%$        & $\pm 0.08\%$        & $\pm 0.07\%$         & $\pm 0.05\%$          & $\pm 0.13\%$      \\ 
\hline
Statistical     
       & $\pm 1.28\%$       & $\pm 0.35\%$        & $\pm 0.32\%$        & $\pm 0.33\%$        & $\pm 0.41\%$          & $\pm 0.88\%$      \\ 
\hline \hline
\multicolumn{7}{|c|}{$\mu D^*$ channel}  \\
\hline
Bin width            & $0.06\%$    & $0.03\%$    & $0.02\%$    & $0.02\%$    & $0.01\%$    & $0.08\%$      \\ 

Fit limits           & $0.05\%$    & $0.01\%$    & $0.01\%$    & $0.01\%$    & $0.06\%$    & $0.06\%$      \\ 

Magnet weighting     & $0.01\%$    & $0.01\%$    & $0.00\%$    & $0.00\%$    & $0.00\%$    & $0.01\%$      \\ 

Signal model         & $0.01\%$    & $0.01\%$    & $0.05\%$    & $0.06\%$    & $0.03\%$    & $0.03\%$      \\    

Background model     & $0.13\%$    & $0.01\%$    & $0.00\%$    & $0.01\%$    & $0.07\%$    & $0.01\%$      \\ 

M($D^{0}$) cut       & $0.01\%$    & $0.01\%$    & $0.01\%$    & $0.01\%$    & $0.02\%$    & $0.02\%$      \\ 
\hline
Combined systematic  & $\pm 0.13\%$    & $\pm 0.04\%$    & $\pm 0.05\%$    & $\pm 0.07\%$    & $\pm 0.08\%$    & $\pm 0.09\%$      \\ 
\hline
Statistical          & $\pm 0.67\%$   & $\pm 0.30\%$   & $\pm 0.30\%$   & $\pm 0.33\%$   & $\pm 0.44\%$   & $\pm 0.99\%$     \\ 
\hline \hline
\end{tabular}
\end{center}
\end{table*}
\renewcommand\arraystretch{1.0}


\section{\label{sec:abkg}Accounting for Detector Asymmetries}

Both channels used in this measurement are reconstructed from the final state particles 
$\mu^{\pm} K^{\pm} \pi^{\mp} \pi^{\mp}$. In relating the measured raw asymmetry to the 
physical asymmetry under investigation, the effects of possible charge asymmetries in particle 
reconstruction must be considered. Neglecting asymmetries of second order or higher, 
the background asymmetry simplifies to:
\begin{eqnarray}
A_{\text{BG}} =  a^{\mu} + a^{K} - 2 a^{\pi}. \label{eq:abkg5}
\end{eqnarray}
where the asymmetries $a^X$ are defined as the difference in reconstruction efficiency 
$\varepsilon$ for the positively and negatively charged particles:
\begin{eqnarray}
a^{X}           &\equiv& \frac{ \varepsilon^{X^+} - \varepsilon^{X^-} } { \varepsilon^{X^+} + \varepsilon^{X^-} }. \label{eq:abkg4}
\end{eqnarray}

\subsection{\label{sec:kaon}Kaon Asymmetry}

By far the largest background asymmetry to be taken into account is due to differences 
in the behavior of positive and negative kaons as they traverse the detector. 
Negative kaons can interact with matter in the tracking system to produce hyperons, 
while there is no equivelent interaction for positive kaons. 
As a result, the mean path length for positive kaons is longer, 
the reconstruction efficiency is higher, and the kaon asymmetry $a^K$ is positive. 

The kaon asymmetry is measured using a dedicated sample of $K^{*0} \to K^+\pi^-$ decays,
based on the technique described in Ref.~\cite{dimuon2010}. 
The $K^+\pi^-$ and $K^-\pi^+$ signal yields are extracted by fitting the charge-specific 
$M(K^{\pm}\pi^{\mp})$ distributions, and the asymmetry is determined by dividing the difference 
by the sum.
The track selection criteria are the same as those required for the $\mu D^{(*)}$ signal channels,
and all events must contain a muon passing the selections described in Section~\ref{sec:esel}.
Since the $K^{*0}$ channel includes a final state pion, of opposite charge to the kaon, 
the correction will also absorb any tracking asymmetry affecting pion reconstruction.
One of the two $a^{\pi}$ terms in Eq.~(\ref{eq:abkg5}) is eliminated as a result. 

\begin{figure}[t]
        \centering
		\includegraphics[width=\columnwidth]{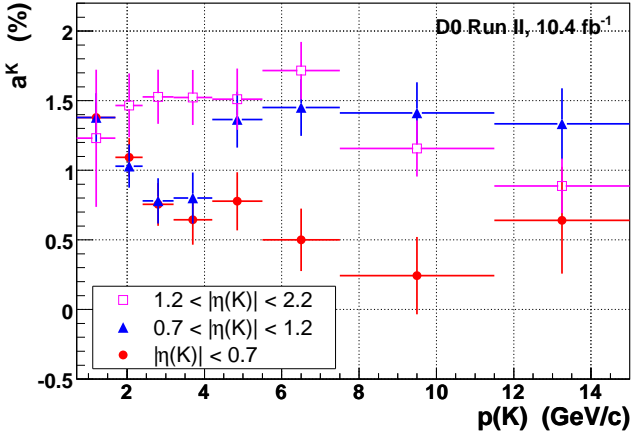}
        \caption[]{Kaon asymmetry as a function of kaon momentum, for three regions 
          of absolute pseudorapidity, as extracted from the $K^{*0} \to K^+\pi^-$ channel.}
\label{fig:kaonasym}
\end{figure}

As expected, an overall positive kaon asymmetry is observed, of approximately 1\% in this channel. 
A strong dependence on kaon momentum and absolute pseudorapidity is found, and hence the final kaon 
asymmetry correction to be applied in Eq.~(\ref{eq:abkg5}) is determined by the weighted average 
of $a^K[p(K),|\eta(K)|]$ over the $p(K)$ and $|\eta(K)|$ distributions in the signal events:
\begin{eqnarray}
a^K =  \sum_{i=1}^{24} a^K_i \cdot \left ( \frac{N^{K}_i}{N} \right ) \label{eq:ak1},
\end{eqnarray}
where the sum is over eight bins in kaon momentum multiplied by three bins in absolute pseudorapidity, 
and $N$ is the total number of signal candidates over all bins. 
The kaon asymmetry as a function of momentum is shown in Fig.~\ref{fig:kaonasym} 
for each of the three $|\eta(K)|$ regions, and the values and bin definitions are 
listed in Table~\ref{tab:kaonasym}. A relative systematic uncertainty of $5\%$ is 
assigned to each bin to account for possible variations in the yield when different 
models are used to fit the signal and backgrounds in the $K^{*0}$ mass distribution.

\renewcommand\arraystretch{1.2}
\begin{table}[t]
\caption[]
{\label{tab:kaonasym}Kaon charge asymmetries in bins of $p(K)$, as extracted from 
the $K^{*0} \to K^+\pi^-$ channel, for each of the three regions in absolute pseudorapidity:
central ($|\eta(K)| < 0.7$); mid-range ($0.7 \leq |\eta(K)| < 1.2$); and forward ($1.2 \leq |\eta(K)| < 2.2$).}
\begin{center}
\begin{tabular}{|ccc|rcl|rcl|rcl|}
\hline \hline
\multicolumn{3}{|c|}{$p(K)$} & \multicolumn{9}{c|}{$a^K$ (\%)}   \\ 
\cline{4-12}
\multicolumn{3}{|c|}{range}  & \multicolumn{3}{c|}{Central}   & \multicolumn{3}{c|}{Mid-Range}  & \multicolumn{3}{c|}{Forward}  \\
\hline \hline
0.7     & -- &   1.7    &  1.38  &  $\pm$ & $0.11$  &  1.38  &  $\pm$ & $0.18$  &  1.23  &  $\pm$ & $0.49$  
\\ \hline
1.7     & -- &   2.4    &  1.09  &  $\pm$ & $0.14$  &  1.03  &  $\pm$ & $0.16$  &  1.47  &  $\pm$ & $0.23$
\\ \hline
2.4     & -- &   3.2    &  0.76  &  $\pm$ & $0.15$  &  0.78  &  $\pm$ & $0.16$  &  1.53  &  $\pm$ & $0.20$
\\ \hline
3.2     & -- &   4.2    &  0.65  &  $\pm$ & $0.18$  &  0.80  &  $\pm$ & $0.18$  &  1.52  &  $\pm$ & $0.20$
\\ \hline
4.2     & -- &   5.5    &  0.78  &  $\pm$ & $0.21$  &  1.36  &  $\pm$ & $0.20$  &  1.51  &  $\pm$ & $0.22$
\\ \hline
5.5     & -- &   7.5    &  0.50  &  $\pm$ & $0.22$  &  1.45  &  $\pm$ & $0.20$  &  1.72  &  $\pm$ & $0.21$
\\ \hline
7.5     & -- &   11.5   &  0.24  &  $\pm$ & $0.28$  &  1.41  &  $\pm$ & $0.22$  &  1.16  &  $\pm$ & $0.20$
\\ \hline
\multicolumn{3}{|c|}{$\geq 11.5$} &  0.64  &  $\pm$ & $0.38$  &  1.33  &  $\pm$ & $0.26$  &  0.89  &  $\pm$ & $0.20$
\\ \hline \hline
\end{tabular}
\end{center}
\end{table}
\renewcommand\arraystretch{1.0}

The kaon momentum distributions for each channel, within each $|\eta(K)|$ region, and for 
each VPDL($B^0$) bin, are determined by fitting the appropriate invariant mass distribution 
in each of the eight kaon momentum bins, using the same parametrizations $F_{\text{sum}}$ 
as descibed in Section~\ref{sec:raw}. 
Following studies over a range of fit variations, a relative systematic uncertainty of 
3\% (0.5\%) is assigned on all $\mu D$ ($\mu D^*$) yields. 

The final corrections for each VPDL bin in both channels are presented in 
Table~\ref{tab:adet_both}. The kaon corrections for the $\mu D^*$ channel are slightly smaller 
than for the $\mu D$ channel, due to different kaon kinematics in the two decay topologies.

\subsection{\label{sec:trk}Track Asymmetry}

Unlike kaons, positive and negative pions have almost identical interaction cross sections in matter.
Any possible asymmetry will be dominated by effects from track detection and reconstruction,
which should be removed to first order by the magnet polarity weighting.

\begin{figure}[h]
        \centering
		\includegraphics[width=\columnwidth]{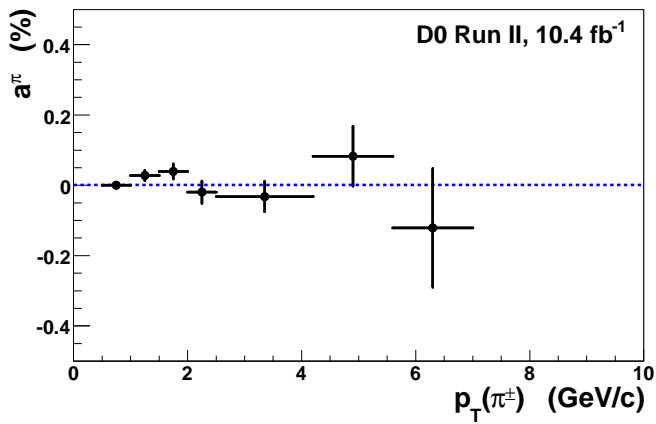}
        \caption[]{Relative track reconstruction asymmetry as a function of track 
transverse momentum, as measured from $K^0_S \to \pi^+ \pi^-$ decays. 
The absolute scale is chosen to give zero asymmetry for the first bin, 
since this channel is only sensitive to the variations in asymmetry between bins.}
\label{fig:trackasym_ks}
\end{figure}

The transverse momentum dependence of any residual tracking asymmetry is studied in
$K^0_S \to \pi^+ \pi^-$ decays. This channel can only be observed if a pair of oppositely 
charged pions is reconstructed, hence it is insensitive to the absolute asymmetry, 
and the overall scale is arbitrarily fixed by setting the asymmetry in the lowest $p_T$ bin to zero.
The relative asymmetry as a function of $p_T$ is determined by extracting the $K_S^0$ yields in 
bins of [$p_T(\pi^+)$,$p_T(\pi^-)$], and following the method described in Ref.~\cite{dimuon2010}, 
except that $K^0_S \to \pi^+ \pi^-$ decays are used instead of $J/\psi \to \mu^+ \mu^-$.
As shown in Fig.~\ref{fig:trackasym_ks}, no evidence of any track $p_T$ dependence is observed, 
over the range $0.5$--$7$~GeV/$c$, within an uncertainty of $\pm 0.05\%$. 
As a result, any residual tracking asymmetry will cancel to first order in the reconstruction of the pion 
and oppositely-charged muon, which remain to be taken into account after applying the kaon asymmetry correction.
There are insufficient statistics in the $K^0_S \to \pi^+ \pi^-$ channel to extend to higher transverse momenta. 
However, this momentum region contains the majority of $\mu D^{(*)}$ signal candidates.

A second dedicated channel $K^{\pm *} \to K^0_S \pi^{\pm}$ is used to measure the absolute residual 
track asymmetry. The $K^0_S \pi^{\pm}$ yields for each pion charge are extracted by fitting the 
$M(K^0_S \pi^{\pm})$ invariant mass distributions, and the asymmetry calculated from the sum and 
difference of these yields. No significant asymmetry is found in this study, 
which is consistent with the findings of previous studies~\cite{dimuon2010}. 
As such, no correction is applied to the asymmetry to account for the effects of track reconstruction,
and $a^{\pi}$ in Eq.~(\ref{eq:abkg5}) is assigned to be zero.
We allocate a systematic uncertainty of $\pm 0.05\%$ to account for the limited precision of this asymmetry measurement.

\subsection{\label{sec:muon}Muon Asymmetry}

\begin{figure}[t]
        \centering
		\includegraphics[width=\columnwidth]{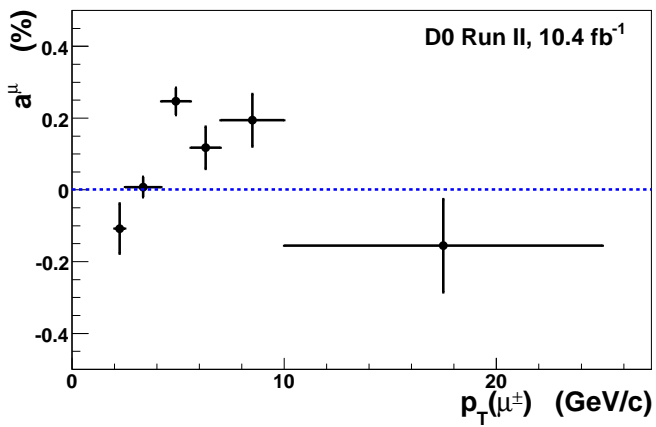}
        \caption[]{Muon asymmetry as a function of muon transverse momentum, 
          as extracted from a dedicated sample of $J/\psi \to \mu^+\mu^-$ decays.}
\label{fig:muonasym}
\end{figure}

The residual charge asymmetry for muon identification is measured using $J/\psi \to \mu^+\mu^-$ decays, 
using the technique developed in Ref.~\cite{dimuon2010}.
A small but significant asymmetry is observed, with a sizeable dependence on the muon 
transverse momentum, as shown in Fig.~\ref{fig:muonasym}. 
The corresponding correction $a^{\mu}$ to be applied to the raw asymmetry is 
extracted using the same method as for the kaon asymmetry, by performing a weighted 
average of the muon asymmetry over bins of $p_T(\mu)$, analogous to Eq.~(\ref{eq:ak1}). 

The final muon asymmetry corrections for each VPDL bin and both channels are 
summarized in Table~\ref{tab:adet_both}.
To account for possible systematic uncertainties, the entire procedure of 
measuring muon asymmetries and convoluting with the transverse momentum distributions 
is repeated with several variations to the method, and the corresponding changes 
in the final measured muon asymmetry in each bin are used to assign a systematic uncertainty.
The variations include changing the mass binning of the $M(\mu^+\mu^-)$ distributions,
changing the fitting function used to extract the $J/\psi$ yields, changing the $p_T$ binning scheme, 
including an absolute pseudorapidity dependence,
and using an alternative method to determine the polarity-based event weights. 

\renewcommand\arraystretch{1.1}
\begin{table*}[t]
\caption[]
{\label{tab:adet_both}The background asymmetries from kaon and muon reconstruction, in bins of VPDL($B^0$), for both signal channels. Also shown are the raw asymmetries $A$, and the combined background asymmetries $A_{\text{BG}} = a^{K} + a^{\mu} - 2 a^{\pi}$, where the final term contributes no net asymmetry but a systematic uncertainty of $\pm 0.05\%$. In each case, the upper uncertainty is statistical, the lower systematic.}
\begin{center}
\begin{tabular}{|l||rcl|rcl|rcl|rcl|rcl|rcl|}
\hline \hline
                           & \multicolumn{3}{c|}{Bin 1} &  \multicolumn{3}{c|}{Bin 2}  &  \multicolumn{3}{c|}{Bin 3}
                             & \multicolumn{3}{c|}{Bin 4} &  \multicolumn{3}{c|}{Bin 5}  &  \multicolumn{3}{c|}{Bin 6} \\

       & \multicolumn{3}{c|}{$-0.10$ -- $0.00$~cm} & \multicolumn{3}{c|}{$0.00$ -- $0.02$~cm}   & \multicolumn{3}{c|}{$0.02$ -- $0.05$~cm}
       & \multicolumn{3}{c|}{$ 0.05$ -- $0.10$~cm} & \multicolumn{3}{c|}{$0.10$ -- $0.20$~cm}   & \multicolumn{3}{c|}{$0.20$ -- $0.60$~cm}  \\ 
\hline \hline
\multicolumn{19}{|c|}{$\mu D$ channel}  \\
\hline
\multirow{2}{*}{$A$ (\%)}                   & 2.70  & $\pm$ & 1.28      & 1.02  & $\pm$ & 0.35     & 1.16  & $\pm$ & 0.32   
                                            & 1.50  & $\pm$ & 0.33      & 1.48  & $\pm$ & 0.41     & 1.20  & $\pm$ & 0.88      \\ 
                                            &       & $\pm$ & 0.19      &       & $\pm$ & 0.07     &       & $\pm$ & 0.08   
                                            &       & $\pm$ & 0.07      &       & $\pm$ & 0.05     &       & $\pm$ & 0.13      \\ 
\hline
\multirow{2}{*}{$a^{K}$ (\%)}               & 1.128 & $\pm$ & 0.041     & 1.124 & $\pm$ & 0.040    & 1.141 & $\pm$ & 0.040   
                                            & 1.147 & $\pm$ & 0.040     & 1.157 & $\pm$ & 0.040    & 1.157 & $\pm$ & 0.040     \\ 
                                            &       & $\pm$ & 0.014     &       & $\pm$ & 0.014    &       & $\pm$ & 0.014   
                                            &       & $\pm$ & 0.014     &       & $\pm$ & 0.015    &       & $\pm$ & 0.014     \\ 
\hline
\multirow{2}{*}{$a^{\mu}$ (\%)}             & 0.102 & $\pm$ & 0.025     & 0.105 & $\pm$ & 0.027    & 0.107 & $\pm$ & 0.029   
                                            & 0.107 & $\pm$ & 0.029     & 0.108 & $\pm$ & 0.028    & 0.108 & $\pm$ & 0.028     \\ 
                                            &       & $\pm$ & 0.008     &       & $\pm$ & 0.009    &       & $\pm$ & 0.012   
                                            &       & $\pm$ & 0.013     &       & $\pm$ & 0.011    &       & $\pm$ & 0.009     \\ 
\hline
\multirow{2}{*}{$A_{\text{BG}}$ (\%)}       & 1.230 & $\pm$ & 0.048     & 1.229 & $\pm$ & 0.048    & 1.248 & $\pm$ & 0.049   
                                            & 1.254 & $\pm$ & 0.049     & 1.265 & $\pm$ & 0.049    & 1.265 & $\pm$ & 0.049     \\ 
                                            &       & $\pm$ & 0.053     &       & $\pm$ & 0.053    &       & $\pm$ & 0.053   
                                            &       & $\pm$ & 0.054     &       & $\pm$ & 0.053    &       & $\pm$ & 0.053     \\ 
\hline \hline
\multicolumn{19}{|c|}{$\mu D^*$ channel}  \\
\hline
\multirow{2}{*}{$A$ (\%)}                   & 1.82  & $\pm$ & 0.67      & 1.10  & $\pm$ & 0.30     & 0.94  & $\pm$ & 0.30   
                                            & 1.38  & $\pm$ & 0.33      & 2.11  & $\pm$ & 0.44     & 0.55  & $\pm$ & 0.99      \\ 
                                            &       & $\pm$ & 0.13      &       & $\pm$ & 0.04     &       & $\pm$ & 0.05   
                                            &       & $\pm$ & 0.07      &       & $\pm$ & 0.08     &       & $\pm$ & 0.09      \\ 
\hline
\multirow{2}{*}{$a^{K}$ (\%)}               & 1.089 & $\pm$ & 0.047     & 1.078 & $\pm$ & 0.052    & 1.078 & $\pm$ & 0.050   
                                            & 1.085 & $\pm$ & 0.050     & 1.086 & $\pm$ & 0.049    & 1.098 & $\pm$ & 0.050     \\ 
                                            &       & $\pm$ & 0.013     &       & $\pm$ & 0.014    &       & $\pm$ & 0.014   
                                            &       & $\pm$ & 0.014     &       & $\pm$ & 0.014    &       & $\pm$ & 0.014     \\ 
\hline
\multirow{2}{*}{$a^{\mu}$ (\%)}             & 0.097 & $\pm$ & 0.027     & 0.098 & $\pm$ & 0.031    & 0.101 & $\pm$ & 0.033   
                                            & 0.101 & $\pm$ & 0.033     & 0.101 & $\pm$ & 0.033    & 0.101 & $\pm$ & 0.031     \\ 
                                            &       & $\pm$ & 0.012     &       & $\pm$ & 0.022    &       & $\pm$ & 0.023   
                                            &       & $\pm$ & 0.022     &       & $\pm$ & 0.020    &       & $\pm$ & 0.016     \\ 
\hline
\multirow{2}{*}{$A_{\text{BG}}$ (\%)}       & 1.186 & $\pm$ & 0.054     & 1.176 & $\pm$ & 0.061    & 1.179 & $\pm$ & 0.060   
                                            & 1.186 & $\pm$ & 0.060     & 1.187 & $\pm$ & 0.059    & 1.199 & $\pm$ & 0.059     \\ 
                                            &       & $\pm$ & 0.053     &       & $\pm$ & 0.056    &       & $\pm$ & 0.057   
                                            &       & $\pm$ & 0.056     &       & $\pm$ & 0.056    &       & $\pm$ & 0.054     \\ 
\hline \hline
\end{tabular}
\end{center}
\end{table*}
\renewcommand\arraystretch{1.0}


\section{\label{sec:adil}Sample Composition: Dilution from Symmetric Processes}

Not all $\mu D^{(*)}$ combinations originate from the decay of oscillated $B^0$ mesons. 
Alternative charge symmetric sources will contribute only to the denominator in the raw asymmetry extraction,
and hence dilute any physical asymmetry $a^d_{\text{sl}}$.

In general, mesons containing a charm quark can be produced in many different ways, 
which we divide into five categories for the purposes of this measurement:
\begin{enumerate}
\item direct hadroniszation from an initial $c (\bar{c})$ quark, here denoted as `prompt'; 
\item as a product of $B^0$ meson decay;
\item as a product of $B^{\pm}$ meson decay;
\item as a product of $B_s^0$ meson decay; and
\item as a product of a $b$ baryon decay.
\end{enumerate} 
The contribution from $b$ baryons is found to be negligible, using the technique described below,
and will no longer be considered. 
This scheme includes possible intermediate excited resonances of both $B$ and $D$ mesons, 
for example processes such as $c \to D^{*0}X \to D^{+}X^{\prime}$ or those with higher excitations. 
For both neutral $B$ meson sources, there may be mixing via box diagrams prior to decay, 
so sources 2 and 4 in the above list can be subdivided into `mixed' and `direct' decays.

The total fraction of signal events coming from $B^0$ meson decays is determined using 
inclusive MC simulations in which the only requirement at the generator level is the presence 
of the appropriate $D^{(*)\mp}$ decay channel, and the presence of a muon (of any charge). 
The samples include non-primary gluon splitting into heavy flavor $b\bar{b}$ and $c\bar{c}$ pairs,
in addition to pair production from flavor excitation and flavor creation mechanisms.
The generated events are passed through the full simulation chain, and then processed 
by the same reconstruction and selection algorithms as used to select events from 
real data for the two signal channels.

At the reconstruction level, the final state tracks must correspond to the true 
kaons and pions from the $D^{(*)\mp}$ decay. 
The result for each channel is a sample of $D^{(*)\mp}$ events, with an accompanying 
reconstructed muon, in which the decay chain can be investigated in detail to extract the 
parentage information.
The candidates are weighted according to their true decay time, to ensure that the $B$ 
meson lifetimes match the current world average measurements, with the uncertainties on 
these lifetimes taken into account when assigning systematic uncertainties.

Table~\ref{tab:fbosc} lists the resulting fractions of $\mu D^{(*)}$ candidates from each source 
in both signal channels, as a function of the reconstructed VPDL($B^0$). 
In general, the $B^0$ fraction is approximately 80--90\%, except in the first (negative VPDL) bin 
in which the prompt contribution reduces this to around $60\%$. 
The $B^{\pm}$ contribution is small but significant, building from approximately 5--18\% 
(6--14\%) in the $\mu D^{(*)}$ case as the VPDL increases. 
This graduation is due to the longer lifetime of the $B^{\pm}$ meson relative to $B^0$.

\renewcommand\arraystretch{1.1}
\begin{table*}[ht]
\caption[]
{\label{tab:fbosc}Fraction of $D^{(*)\pm}$ candidates arising from each source as determined from MC simulation, 
for both channels, and in bins of VPDL($B^0$). For each channel, the upper row lists the central values 
and statistical uncertainties (from limited sample size in simulation), while the lower row lists the 
systematic uncertainties, only determined for the final $F_{B^0}^{\text{osc}}$ values.}
\begin{center}
\begin{tabular}{|l||rcl|rcl|rcl|rcl|rcl|rcl|}
\hline \hline
                           & \multicolumn{3}{c|}{Bin 1} &  \multicolumn{3}{c|}{Bin 2}  &  \multicolumn{3}{c|}{Bin 3}
                             & \multicolumn{3}{c|}{Bin 4} &  \multicolumn{3}{c|}{Bin 5}  &  \multicolumn{3}{c|}{Bin 6} \\

       & \multicolumn{3}{c|}{$-0.10$ -- $0.00$~cm} & \multicolumn{3}{c|}{$0.00$ -- $0.02$~cm}   & \multicolumn{3}{c|}{$0.02$ -- $0.05$~cm}
       & \multicolumn{3}{c|}{$ 0.05$ -- $0.10$~cm} & \multicolumn{3}{c|}{$0.10$ -- $0.20$~cm}   & \multicolumn{3}{c|}{$0.20$ -- $0.60$~cm}     \\ 
\hline \hline
\multicolumn{19}{|c|}{$\mu D$ channel}  \\
\hline
$F(\bar{c} \to \mu D X)$                      & 0.361 & $\pm$ & 0.011          & 0.069 & $\pm$ & 0.003          & 0.003 & $\pm$ & 0.000   
                                                & 0.000 & $\pm$ & 0.000          & 0.000 & $\pm$ & 0.000          & 0.000 & $\pm$ & 0.000  \\

$F(B^0_s \to \mu D X)$                        & 0.019 & $\pm$ & 0.003          & 0.019 & $\pm$ & 0.001          & 0.029 & $\pm$ & 0.001   
                                                & 0.027 & $\pm$ & 0.001          & 0.030 & $\pm$ & 0.002          & 0.032 & $\pm$ & 0.004  \\

$F(B^{\pm} \to \mu D X)$                      & 0.052 & $\pm$ & 0.005          & 0.075 & $\pm$ & 0.003          & 0.101 & $\pm$ & 0.003   
                                                & 0.118 & $\pm$ & 0.003          & 0.141 & $\pm$ & 0.004          & 0.186 & $\pm$ & 0.008  \\

$F(B^0 \to \mu D X)$                          & 0.569 & $\pm$ & 0.011          & 0.837 & $\pm$ & 0.004          & 0.868 & $\pm$ & 0.003   
                                                & 0.854 & $\pm$ & 0.003          & 0.829 & $\pm$ & 0.004          & 0.781 & $\pm$ & 0.009  \\
\hline
\multirow{2}{*}{$F_{B^0}^{\text{osc}}$}     & 0.018 & $\pm$ & 0.003          & 0.009 & $\pm$ & 0.001          & 0.057 & $\pm$ & 0.002   
                                                & 0.208 & $\pm$ & 0.003          & 0.520 & $\pm$ & 0.005          & 0.658 & $\pm$ & 0.010   \\ 
                                                &       & $\pm$ & 0.001          &       & $\pm$ & 0.000          &       & $\pm$ & 0.001   
                                                &       & $\pm$ & 0.005          &       & $\pm$ & 0.011          &       & $\pm$ & 0.017   \\ 
\hline \hline
\multicolumn{19}{|c|}{$\mu D^*$ channel}  \\
\hline
$F(\bar{c} \to \mu D^* X)$                   & 0.373 & $\pm$ & 0.010          & 0.082 & $\pm$ & 0.003          & 0.005 & $\pm$ & 0.001   
                                                & 0.000 & $\pm$ & 0.000          & 0.001 & $\pm$ & 0.000          & 0.000 & $\pm$ & 0.000  \\

$F(B^0_s \to \mu D^* X)$                     & 0.009 & $\pm$ & 0.002          & 0.011 & $\pm$ & 0.001          & 0.014 & $\pm$ & 0.001   
                                                & 0.013 & $\pm$ & 0.001          & 0.016 & $\pm$ & 0.002          & 0.017 & $\pm$ & 0.005  \\

$F(B^{\pm} \to \mu D^* X)$                   & 0.058 & $\pm$ & 0.005          & 0.073 & $\pm$ & 0.003          & 0.080 & $\pm$ & 0.003   
                                                & 0.083 & $\pm$ & 0.003          & 0.104 & $\pm$ & 0.005          & 0.146 & $\pm$ & 0.013  \\

$F(B^0 \to \mu D^* X)$                       & 0.560 & $\pm$ & 0.010          & 0.835 & $\pm$ & 0.004          & 0.901 & $\pm$ & 0.003   
                                                & 0.903 & $\pm$ & 0.004          & 0.880 & $\pm$ & 0.005          & 0.836 & $\pm$ & 0.013  \\
\hline
\multirow{2}{*}{$F_{B^0}^{\text{osc}}$}    & 0.013 & $\pm$ & 0.002          & 0.010 & $\pm$ & 0.001          & 0.061 & $\pm$ & 0.003
                                                & 0.231 & $\pm$ & 0.005          & 0.570 & $\pm$ & 0.008          & 0.713 & $\pm$ & 0.016   \\
                                                &       & $\pm$ & 0.001          &       & $\pm$ & 0.000          &       & $\pm$ & 0.002   
                                                &       & $\pm$ & 0.003          &       & $\pm$ & 0.007          &       & $\pm$ & 0.008       \\ 
\hline \hline
\end{tabular}
\end{center}
\end{table*}
\renewcommand\arraystretch{1.0}

For all bins, the $B^0_s$ fraction is very small, at approximately 1--3\%. 
We correct for the possible contribution from the semileptonic mixing asymmetry in 
$B^0_s$ mesons, $a^s_{\text{sl}}$, on a VPDL bin-by-bin basis, by extending Eq.~(\ref{eq:main}) 
to include the non-zero $B^0_s$ fraction:
\begin{eqnarray}
a^d_{\text{sl}} &=& \frac{A - A_{\text{BG}} - F_{B_s^0}^{\text{osc}} \cdot a^s_{\text{sl}}}{F_{B^0}^{\text{osc}}}. \label{eq:main2}
\end{eqnarray}
Here $F_{B_s^0}^{\text{osc}} = F(B^0_s \to \mu D^{(*)}X) \cdot \chi_s$ is the fractional 
contribution of oscillated $B^0_s$ mesons in the sample, where $\chi_s = 0.499292 \pm 0.000016$
is the integrated mixing probability~\cite{pdg}.
The parameter $a^s_{\text{sl}}$ is assigned the world average value, $[-1.05 \pm 0.64]\%$~\cite{pdg}. 
The uncertainty on this quantity is taken into account when determining the 
systematic uncertainty on the final measurement.

The fraction of $B^0$ mesons that oscillate into their antiparticle prior to decay
is determined by applying a weight $W_j^{\text{mix}}$ to all $B^0 \to \mu D^{(*)}$ events based on the true decay time 
($t_j$) of the $B^0$ meson:
\begin{equation}
W_j^{\text{mix}} = \frac{1}{2}[1 - \text{cos}(\Delta M_d \cdot t_j)],
\end{equation}
where $\Delta M_d$ is the mass difference of the heavy and light eigenstates in the 
$B^0$ system, assigned to be the world-average value $0.507 \pm 0.004$~ps$^{-1}$~\cite{pdg}, 
with the precision taken into account when assigning systematic uncertainties. 
The resulting fractions $F_{B^0}^{\text{osc}}$, are defined as the sum of these 
mixing weights divided by the total number of events in the MC sample. 
Table~\ref{tab:fbosc} lists the resulting fractions versus VPDL for both channels.

Various sources of systematic uncertainty on $F_{B^0}^{\text{osc}}$ are considered.
The prompt fraction is negligible in VPDL bins 3--6 used in the final $a^d_{\text{sl}}$ measurement; 
therefore, no systematic uncertainties are allocated from this source.

The simulation may not describe the data perfectly. In particular, the simulation doesn't 
account for any effects due to the muon triggers used to collect data.
MC simulations show that the pre-trigger muon transverse momenta distributions from $B^0$ and $B^{\pm}$ decays 
are completely consistent, as expected from the closeness of the meson masses.
On the other hand, there are small differences in the $p_T(\mu)$ distributions for $B_s^0$ decays.
Reweighting events by a trigger acceptance correction leads to a small reduction in the $B_s^0$ fraction,
of order 3\%. Since this source accounts for less than $3\%$ of all $D^{(*)-}$ candidates, 
the effect of this trigger correction on $F_{B^0}^{\text{osc}}$ is tiny, of order 0.001, and is neglected.

The decay branching ratios of $B^0$ mesons into semileptonic final states containing a $D^{\mp}$ ($D^{*\mp}$)
meson are known to around 10\% (5\%) precision~\cite{pdg}. As such, we vary the $B^0$ fractions up and down 
for the two channels by these fractions and assign a systematic uncertainty from this source equal to the 
total variation with respect to the default value.


To account for the uncertainties on the world-average $B$ meson lifetime values, 
we repeat the evaluation of $F_{B^0}^{\text{osc}}$ with the input lifetimes adjusted within their 
uncertainties, and assign a systematic uncertainty equal to the maximum deviation from the 
nominal $F_{B^0}^{\text{osc}}$ value.
Similarly, systematic uncertainties are allocated to account for the limited precision of $\Delta M_d$ and $a^s_{\text{sl}}$.
The breakdown of systematic uncertainties is shown in Table~\ref{tab:fbosc_syst}.
For the final measurement of $a^d_{\text{sl}}$, the uncertainties from the limited 
sample size in simulation are also categorized as systematic, 
not statistical, since they are not related to the size of the data sample.

\renewcommand\arraystretch{1.1}
\begin{table*}[t]
\caption[]
{\label{tab:fbosc_syst}Systematic uncertainties from different sources on the dilution fraction $F_{B^0}^{\text{osc}}$, for both channels, and in bins of VPDL($B^0$).}
\begin{center}
\begin{tabular}{|l||rl|rl|rl|rl|rl|rl|}
\hline \hline
                           & \multicolumn{2}{c|}{Bin 1} &  \multicolumn{2}{c|}{Bin 2}  &  \multicolumn{2}{c|}{Bin 3}
                             & \multicolumn{2}{c|}{Bin 4} &  \multicolumn{2}{c|}{Bin 5}  &  \multicolumn{2}{c|}{Bin 6} \\

       & \multicolumn{2}{c|}{$-0.10$ -- $0.00$~cm} & \multicolumn{2}{c|}{$0.00$ -- $0.02$~cm}   & \multicolumn{2}{c|}{$0.02$ -- $0.05$~cm}
       & \multicolumn{2}{c|}{$ 0.05$ -- $0.10$~cm} & \multicolumn{2}{c|}{$0.10$ -- $0.20$~cm}   & \multicolumn{2}{c|}{$0.20$ -- $0.60$~cm}  \\ 
\hline \hline
\multicolumn{13}{|c|}{$F_{B^0}^{\text{osc}}(\mu D)$} \\
\hline
Branching Ratios                    &  $\pm$ & 0.001          &     $\pm$ & 0.000          &     $\pm$ & 0.001   
                                    &  $\pm$ & 0.004          &     $\pm$ & 0.009          &     $\pm$ & 0.015       \\ 


$B$ meson lifetimes                 &  $\pm$ & 0.000          &     $\pm$ & 0.000          &     $\pm$ & 0.000   
                                    &  $\pm$ & 0.001          &     $\pm$ & 0.003          &     $\pm$ & 0.007       \\ 

$\Delta M_d$                        &  $\pm$ & 0.000          &     $\pm$ & 0.000          &     $\pm$ & 0.001   
                                    &  $\pm$ & 0.003          &     $\pm$ & 0.005          &     $\pm$ & 0.002       \\ 
\hline
Total                               &  $\pm$ & 0.001          &     $\pm$ & 0.000          &     $\pm$ & 0.001   
                                    &  $\pm$ & 0.005          &     $\pm$ & 0.011          &     $\pm$ & 0.017       \\ 
\hline \hline
\multicolumn{13}{|c|}{$F_{B^0}^{\text{osc}}(\mu D^{*})$}  \\
\hline
Branching Ratios                    &  $\pm$ & 0.001          &     $\pm$ & 0.000          &     $\pm$ & 0.001   
                                    &  $\pm$ & 0.001          &     $\pm$ & 0.004          &     $\pm$ & 0.006       \\ 


$B$ meson lifetimes                 &  $\pm$ & 0.000          &     $\pm$ & 0.000          &     $\pm$ & 0.001   
                                    &  $\pm$ & 0.001          &     $\pm$ & 0.003          &     $\pm$ & 0.005       \\ 

$\Delta M_d$                        &  $\pm$ & 0.000          &     $\pm$ & 0.000          &     $\pm$ & 0.001   
                                    &  $\pm$ & 0.003          &     $\pm$ & 0.005          &     $\pm$ & 0.003       \\ 
\hline
Total                               &  $\pm$ & 0.001          &     $\pm$ & 0.000          &     $\pm$ & 0.002   
                                    &  $\pm$ & 0.003          &     $\pm$ & 0.007          &     $\pm$ & 0.008       \\ 
\hline \hline
\end{tabular}
\end{center}
\end{table*}
\renewcommand\arraystretch{1.0}


\section{\label{sec:results}Results}

From the raw asymmetries and detector-related asymmetries listed in Table~\ref{tab:adet_both}, 
and the corresponding dilution fractions $F_{B^0}^{\text{osc}}$ presented in Table~\ref{tab:fbosc}, 
the final value of the semileptonic mixing asymmetry $a^d_{\text{sl}}$ is determined for each VPDL bin 
$i=\{3,4,5,6\}$, and for each channel $j=\{\mu D, \mu D^*\}$:
\begin{eqnarray}
a^d_{\text{sl}}(ij) &=& \frac{A(ij) - a^K(ij) - a^{\mu}(ij) - F_{B_s^0}^{\text{osc}}(ij) \cdot a^s_{\text{sl}}}{F_{B^0}^{\text{osc}}(ij)}.\nonumber \\
\end{eqnarray}
The first two VPDL bins, $i=\{1,2\}$ are not included, as these represent the control region 
in which the expected signal contribution is negligible.

To extract the corresponding uncertainties (both statistical and systematic) care must be 
taken to properly account for all the correlations and constraints on the various inputs to the measurement. In particular:
\begin{itemize}
\item the raw asymmetries for each bin and channel are independent;
\item the kaon asymmetry as a function of [$p(K)$,$|\eta(K)|$] (Fig.~\ref{fig:kaonasym}) 
is 100\% correlated between bins, and between channels;
\item the muon asymmetry as a function of $p_T(\mu)$ (Fig.~\ref{fig:muonasym}) 
is 100\% correlated between bins, and between channels;
\item the asymmetry $a^s_{\text{sl}}$, used to derive the correction for a possible contribution 
from $B^0_s$ mixing, is 100\% correlated between bins, and between channels;
\item the oscillation fractions, $F_{B^0}^{\text{osc}}$, are treated as independent. 
\end{itemize}
While we expect some correlations between bins, and between channels, in the oscillated $B^0$ fractions $F_{B^0}^{\text{osc}}$,
studies indicate that their effect on the final $a^d_{\text{sl}}$ measurement is negligible, justifying their exclusion.

To ensure that all such correlations are taken into account, the final statistical and 
systematic uncertainties on each $a^d_{\text{sl}}$ measurement, and on the combination, are derived 
from 200\ts000 ensemble tests in which all input variables are randomly chosen according to a Gaussian probability 
density function, with an appropriate central value and width, and the distributions of the resulting 
$a^d_{\text{sl}}$ measurements are inspected and fitted. 
This process is performed twice, once with the inputs varied according 
to their statistical uncertainties, and once with the inputs varied according to their systematic uncertainties.

\begin{figure*}[t] 
        \centering
	\subfigure[~$\mu D$ Channel]
        {\includegraphics[width=\columnwidth]{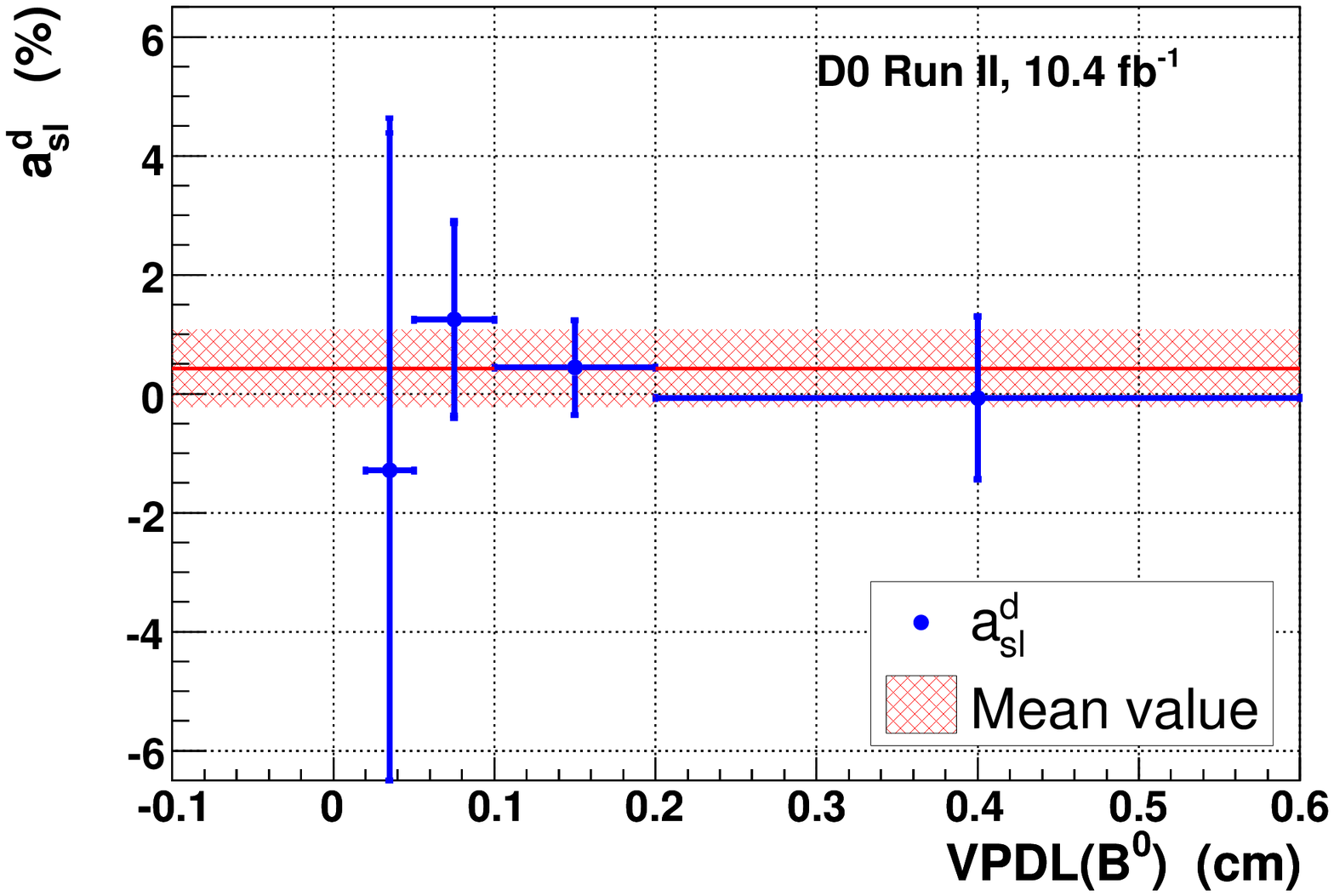}}
        \subfigure[~$\mu D^*$ Channel]
        {\includegraphics[width=\columnwidth]{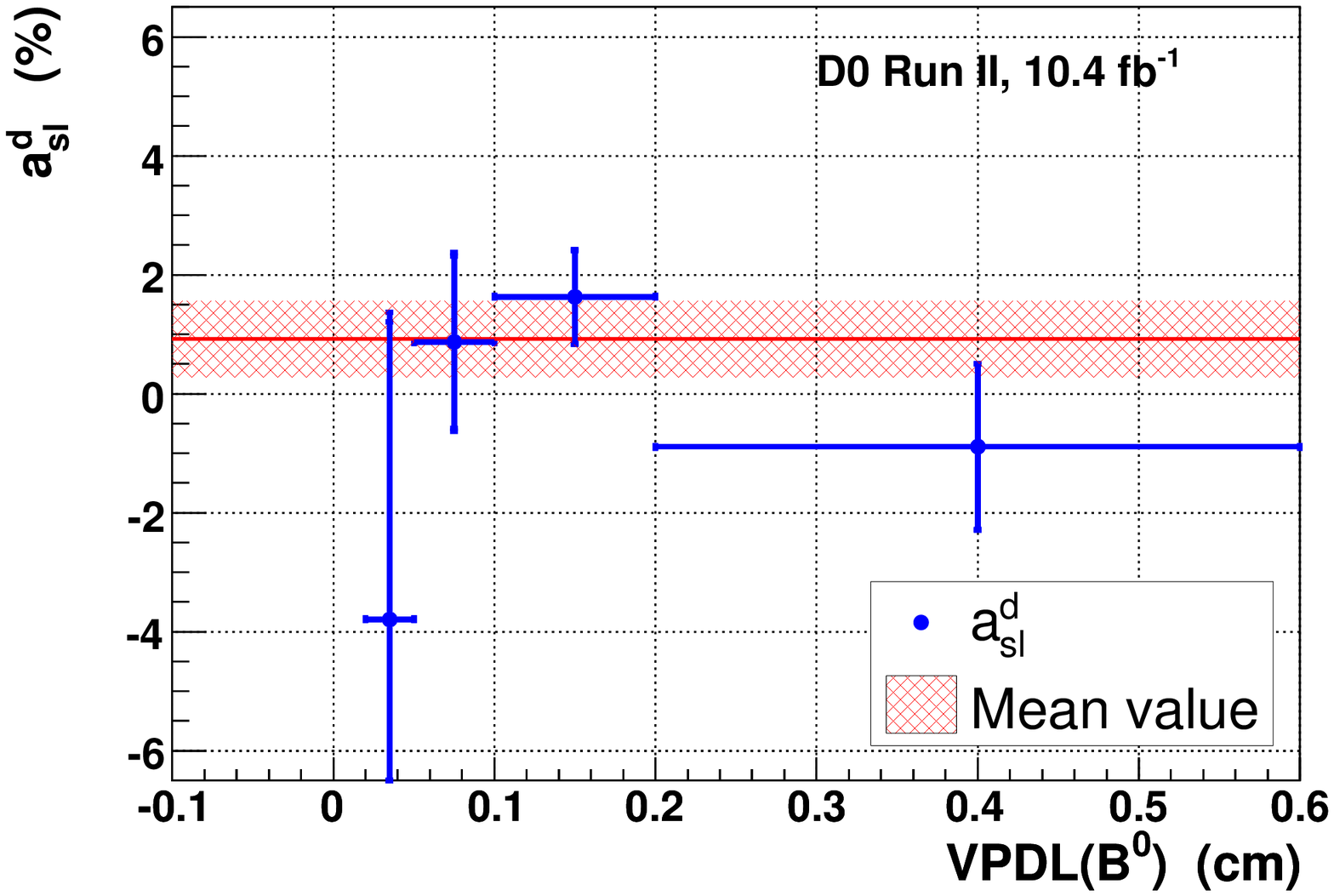}}
        \caption[]{Final measurements of the semileptonic asymmetry $a^d_{\text{sl}}$, in bins of VPDL($B^0$), for both channels. The cross-hatched bands show the mean values (and their total uncertainties) determined for each channel separately.}
\label{fig:adsl_results}
\end{figure*}

Figure~\ref{fig:adsl_results} and Table~\ref{tab:adsl_results} show the individual results for the four signal 
VPDL bins in each channel, with statistical and systematic uncertainties.

\renewcommand\arraystretch{1.1}
\begin{table*}[ht]
\caption[]
{\label{tab:adsl_results}Individual measurements of ($A - A_{\text{BG}}$) 
and $a^d_{\text{sl}}$ in each of the six VPDL($B^0$) bins, for both channels in this analysis. 
For each entry, the upper uncertainty is statistical, the lower systematic, 
with all correlations taken into account.
Also shown are the weights $w(ij)$ used to combine the eight separate measurements, 
normalized to unity.}
\begin{center}
\begin{tabular}{|l||rcl|rcl|rcl|rcl|rcl|rcl|}
\hline \hline
       &  \multicolumn{3}{c|}{Bin 1}   & \multicolumn{3}{c|}{Bin 2} 
       &  \multicolumn{3}{c|}{Bin 3}   & \multicolumn{3}{c|}{Bin 4} & \multicolumn{3}{c|}{Bin 5} &  \multicolumn{3}{c|}{Bin 6}  \\
       &  \multicolumn{3}{c|}{$-0.10$ -- $0.00$~cm}  & \multicolumn{3}{c|}{$ 0.00$ -- $0.02$~cm}  & \multicolumn{3}{c|}{ $0.02$ -- $0.05$~cm}    
       & \multicolumn{3}{c|}{ $0.05$ -- $0.10$~cm}   & \multicolumn{3}{c|}{ $0.10$ -- $0.20$~cm}  & \multicolumn{3}{c|}{ $0.20$ -- $0.60$~cm} \\ 
\hline \hline
\multicolumn{19}{|c|}{$\mu D$ Channel} \\[2pt]
\hline
\multirow{2}{*}{$A - A_{\text{BG}}$ (\%)}                      & 1.48 & $\pm$ & 1.28          & $-0.20$ & $\pm$ & 0.35  
                                   & $-0.07$ & $\pm$ & 0.32    & 0.26  & $\pm$ & 0.33         & 0.23  & $\pm$ & 0.41     & $-0.05$ & $\pm$ & 0.89  \\ 
                                                               &      & $\pm$ & 0.20          &       & $\pm$ & 0.09  
                                   &     & $\pm$ & 0.10        &      & $\pm$ & 0.09          &       & $\pm$ & 0.07     &     & $\pm$ & 0.14     \\ 
\hline
\multirow{2}{*}{$a^d_{\text{sl}}$ (\%)}                        &\multicolumn{6}{c|}{\multirow{3}{*}{Not used}}
                                   & $-1.29$ & $\pm$ & 5.68    & 1.25 & $\pm$ & 1.61          & 0.44 & $\pm$ & 0.79     & -0.07 & $\pm$ & 1.36      \\ 
                                                               &\multicolumn{6}{c|}{}
                                   &         & $\pm$ & 1.69    &      & $\pm$ & 0.43          &      & $\pm$ & 0.14     &     & $\pm$ & 0.21  \\ 
\cline{1-1} \cline{8-19}
weight $w(ij) / \sum_{ij} w(ij)$                                &\multicolumn{6}{c|}{}      
                                   & \multicolumn{3}{c|}{0.006} & \multicolumn{3}{c|}{0.072}  & \multicolumn{3}{c|}{0.309}  & \multicolumn{3}{c|}{0.105}   \\ 
\hline \hline
\multicolumn{19}{|c|}{$\mu D^*$ Channel}  \\[2pt]
\hline
\multirow{2}{*}{$A - A_{\text{BG}}$ (\%)}                      & 0.64 & $\pm$ & 0.67  & $-0.07$ & $\pm$ & 0.31  
                                   & $-0.23$ & $\pm$ & 0.31    & 0.20 & $\pm$ & 0.34  & 0.93 & $\pm$ & 0.44  & $-0.63$ & $\pm$ & 0.99      \\ 
                                                               &      & $\pm$ & 0.14  &      & $\pm$ & 0.07  
                                   &     & $\pm$ & 0.08        &      & $\pm$ & 0.09  &      & $\pm$ & 0.10  &     & $\pm$ & 0.11     \\ 
\hline
\multirow{2}{*}{$a^d_{\text{sl}}$ (\%)}                          &\multicolumn{6}{c|}{\multirow{3}{*}{Not used}}    
                                   & $-3.79$ & $\pm$ & 5.00      & 0.87  & $\pm$ & 1.45         & 1.63 & $\pm$ & 0.78     & $-0.89$ & $\pm$ & 1.39      \\ 
                                                                 &\multicolumn{6}{c|}{}
                                   &     & $\pm$ & 1.27          &      & $\pm$ & 0.39          &      & $\pm$ & 0.17     &     & $\pm$ & 0.15      \\ 
\cline{1-1} \cline{8-19}
weight $w(ij) / \sum_{ij} w(ij)$                                &\multicolumn{6}{c|}{}      
                                   & \multicolumn{3}{c|}{0.007} & \multicolumn{3}{c|}{0.088}    & \multicolumn{3}{c|}{0.311}  & \multicolumn{3}{c|}{0.102}   \\ 
\hline \hline
\end{tabular}
\end{center}
\end{table*}
\renewcommand\arraystretch{1.0}

Once the uncertainties on the individual $a^d_{\text{sl}}$ measurements are established, 
the combination between VPDL bins, and then between channels, is performed. 
For each channel, the combined $a^d_{\text{sl}}$ value is obtained by a weighted average 
of the four individual measurements:
\begin{eqnarray}
a^d_{\text{sl}}(j) & = & \frac{\sum_{i=3}^6 a^d_{\text{sl}}(ij) w(ij)}{\sum_{i=3}^6 w(ij)},
\end{eqnarray}
where the weights $w(ij)$ are the inverse of the sum in quadrature of statistical 
and systematic uncertainties for that measurement:
\begin{eqnarray}
w(ij) & = & \frac{1}{\sigma_{\text{stat}}^2[a^d_{\text{sl}}(ij)] +\sigma_{\text{syst}}^2[a^d_{\text{sl}}(ij)]}.
\end{eqnarray}
The central values and uncertainties for the combinations are again determined by performing 
the full set of 200,000 ensemble tests, with all inputs varied, and examining the effect on the 
final values of $a^d_{\text{sl}}$ from each channel. This procedure yields the following results:
\begin{eqnarray}
a^d_{\text{sl}}(\mu D)   & = & [0.43 \pm 0.63 \text{ (stat.)} \pm 0.16 \text{ (syst.)}]\%,\text{ ~  ~} \\
a^d_{\text{sl}}(\mu D^*) & = & [0.92 \pm 0.62 \text{ (stat.)} \pm 0.16 \text{ (syst.)}]\%.\text{ ~  ~}
\end{eqnarray}
Finally, the combination is extended to give the full weighted average of the two channel-specific
measurements, with full propagation of uncertainties, to yield the final measurement:
\begin{eqnarray}
a^d_{\text{sl}} & = & [0.68 \pm 0.45 \text{ (stat.)} \pm 0.14 \text{ (syst.)}]\%.
\end{eqnarray}
The weights $w(ij)$ used for this combination are presented in Table~\ref{tab:adsl_results}.


\section{\label{sec:checks}Cross-checks}

To test the robustness of the measurement technique, the analysis is repeated with the event samples 
divided into pairs of orthogonal sub-sets, of approximately equal size. 
The raw asymmetries $A$, detector-related background corrections $a^K$ and $a^{\mu}$, 
and oscillation fractions $F_{B^0}^{\text{osc}}$ are redetermined for each sub-sample, 
and the semileptonic mixing asymmetry $a^d_{\text{sl}}$ measured in each case. 
The sub-samples are defined by the following criteria:
\begin{itemize}
\item $\eta(\mu) < 0$ and $\eta(\mu) > 0$;
\item $|\eta(K)| < 0.7$ and $|\eta(K)| > 0.7$;
\item $p(K) < 3.2$~GeV/$c$ and $p(K) > 3.2$~GeV/$c$;
\item a chronological division corresponding to early and late data collection;
\item $\sigma(\text{VPDL}) < 40$~$\mu$m and $\sigma(\text{VPDL}) > 40$~$\mu$m.
\end{itemize}
The results are summarized in Table~\ref{tab:crosschecks}.
In all cases, the measured values of $a^d_{\text{sl}}$ are statistically consistent with each other,
despite some samples having significantly different background corrections.

\renewcommand\arraystretch{1.2}
\begin{table}[t]
\caption[]
{\label{tab:crosschecks}Results of the analysis cross-checks, in which the data are divided into pairs of 
orthogonal and independent subsets, and the measurements of $a^d_{\text{sl}}$ repeated for each sample. 
The uncertainties shown here are the sum in quadrature of the statistical and systematic components.}
\begin{center}
\begin{tabular}{|c|rcl|rcl|rcl|}
\hline \hline
Sub-sample              & \multicolumn{9}{c|}{$a^d_{\text{sl}}$ (\%)}   \\ 
\cline{2-10}
requirement             & \multicolumn{3}{c|}{$\mu D$ channel}   & \multicolumn{3}{c|}{$\mu D^*$ channel}  & \multicolumn{3}{c|}{Comb.}  \\
\hline \hline
Nominal Result                             & 0.43  & $\pm$ & 0.65    & 0.92     & $\pm$ & 0.64    & 0.68     & $\pm$ & 0.47        \\   \hline
$\eta(\mu) < 0$                            & 0.38  & $\pm$ & 0.88    & 0.60     & $\pm$ & 0.88    & 0.49     & $\pm$ & 0.63       \\
$\eta(\mu) > 0$                            & 0.53  & $\pm$ & 0.91    & 1.21     & $\pm$ & 0.88    & 0.88     & $\pm$ & 0.64       \\   \hline
$|\eta(K)| < 0.7$                          & 0.48  & $\pm$ & 0.95    & $-0.39$  & $\pm$ & 1.14    & 0.04     & $\pm$ & 0.77       \\
$|\eta(K)| > 0.7$                          & 0.36  & $\pm$ & 0.85    & 1.17     & $\pm$ & 0.86    & 0.76     & $\pm$ & 0.62       \\   \hline
$p(K) < 3.2$~GeV/$c$                       & 0.02  & $\pm$ & 0.87    & $-0.30$  & $\pm$ & 1.42    & $-0.14$  & $\pm$ & 0.85       \\
$p(K) > 3.2$~GeV/$c$                       & 1.11  & $\pm$ & 0.92    & 1.00     & $\pm$ & 0.79    & 1.05     & $\pm$ & 0.62       \\   \hline
$\sigma(\text{VPDL}) < 40$~$\mu m$         & 0.22  & $\pm$ & 0.84    & 0.18     & $\pm$ & 0.95    & 0.20     & $\pm$ & 0.65       \\
$\sigma(\text{VPDL}) > 40$~$\mu m$         & 0.76  & $\pm$ & 0.95    & 0.98     & $\pm$ & 1.01    & 0.87     & $\pm$ & 0.70       \\   \hline
First half data                            & 0.82  & $\pm$ & 0.89    & 1.39     & $\pm$ & 0.88    & 1.11     & $\pm$ & 0.67       \\
Second half data                           & 0.19  & $\pm$ & 0.86    & $-0.30$  & $\pm$ & 1.02    & $-0.06$  & $\pm$ & 0.68       \\   \hline  \hline
\end{tabular}
\end{center}
\end{table}
\renewcommand\arraystretch{1.0}

In addition, the measurement is repeated using only events that satisfy a single muon trigger. 
This corresponds to approximately 90\% of the total sample. The resulting $a^d_{\text{sl}}$ value for these events
is consistent with the nominal value, taking into account the correlation between the samples.

\begin{figure*}[t] 
        \centering
	\subfigure[~$\mu D$ Channel]
        {\includegraphics[width=\columnwidth]{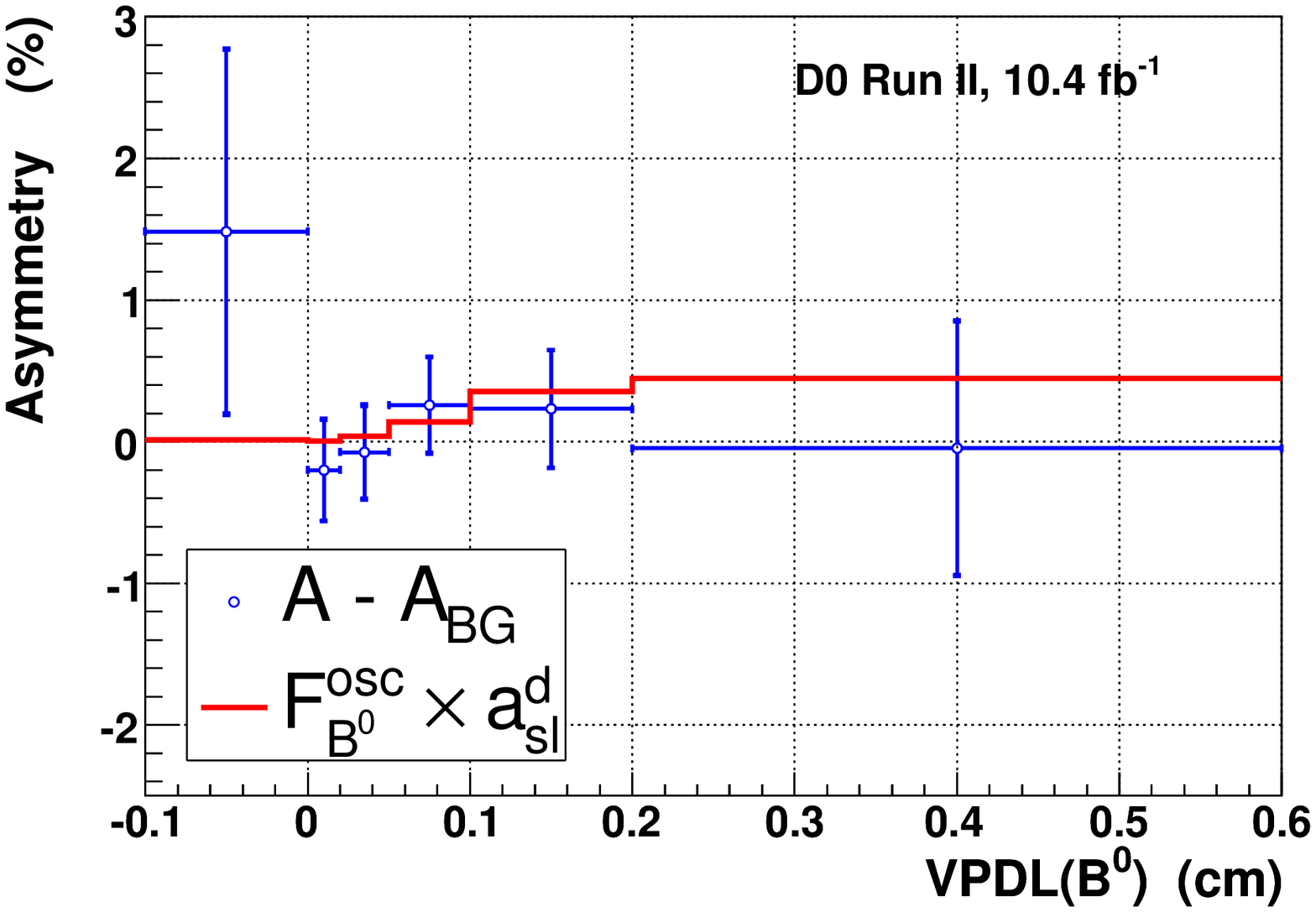}}
        \subfigure[~$\mu D^*$ Channel]
        {\includegraphics[width=\columnwidth]{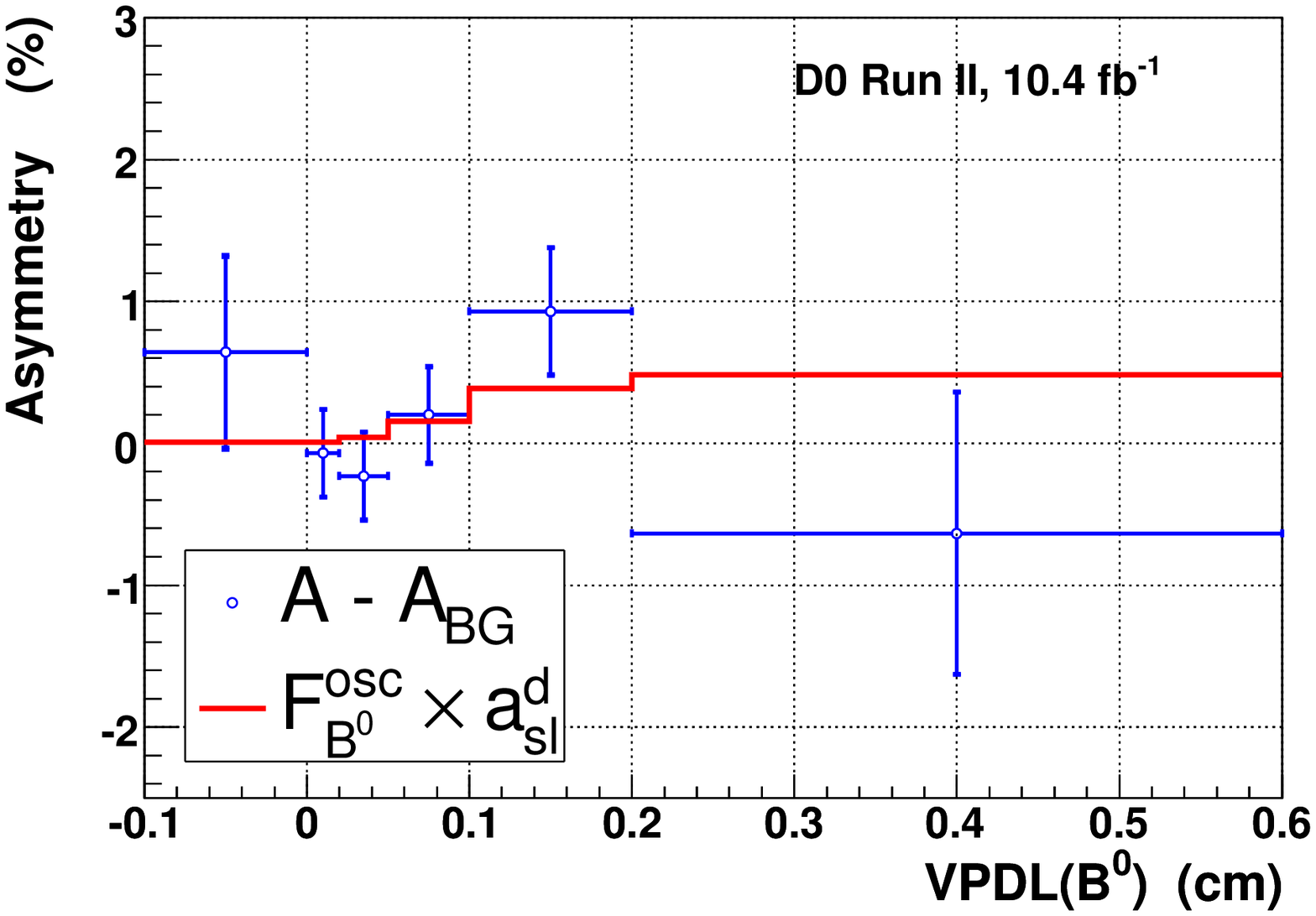}}
        \caption[]{Final measurements of the background corrected asymmetry, in bins of VPDL($B^0$), 
          for both channels. The points show the observed asymmetry, with the solid lines showing 
          $F_{B^0}^{\text{osc}} \cdot a^d_{\text{sl}}$. 
          Any asymmetry caused by mixing should exhibit a characteristic
          turn-on shape as the fraction of oscillated $B^0$ mesons increases.}
\label{fig:aphys_results}
\end{figure*}

The fraction of events from mixed $B^0$ decays, $F_{B^0}^{\text{osc}}$, is a strong function of the 
visible proper decay length of the reconstructed $B^0$ candidate.
Hence any non-zero value of $a^d_{\text{sl}}$ will lead to a VPDL dependence on the 
background subtracted asymmetry $(A - A_{\text{BG}})$ [see Eq.~(\ref{eq:main})]. 
Figure~\ref{fig:aphys_results} shows this dependence for both channels, with the $F_{B^0}^{\text{osc}} \cdot a^d_{\text{sl}}$ 
distribution superimposed on the plot for comparison, using the final $a^d_{\text{sl}}$ measurement from the two channel combination. 
The two distributions are statistically consistent, indicating that the VPDL dependence of the 
observed background-subtracted asymmetry is consistent with the hypothesis that it originates from the mixing of $B^0$ mesons. 
The $\chi^2$ quantifying this agreement between the $(A - A_{\text{BG}})$ and $F_{B^0}^{\text{osc}} \cdot a^d_{\text{sl}}$
distributions is 2.3 (4.5) for the $\mu D^{(*)}$ channel, compared to 2.7 (6.9) under the SM assumption for $a^d_{\text{sl}}$.
For this test, the statistical and systematic uncertainties are combined in quadrature.

The same data can be used to validate the $a^d_{\text{sl}}$ measurement using an alternative method, in which the distribution of 
$(A - A_{\text{BG}})$ versus VPDL($B^0$) is fitted to the function:
\begin{eqnarray}
F(\text{VPDL}) = A_{\text{const}} + F_{B^0}^{\text{osc}}(\text{VPDL}) \cdot a^d_{\text{sl}},
\end{eqnarray}
where $a^d_{\text{sl}}$ and $A_{\text{const}}$ are the two free parameters. The constant asymmetry term allows for a contribution
from possible additional background sources of asymmetry that have not been considered in this analysis.
For this study we neglect any uncertainties on  $F_{B^0}^{\text{osc}}$.
The results are as follows:
\begin{eqnarray}
a^d_{\text{sl}} &=& (0.51 \pm 0.86) \% ~\text{($\mu D$ channel)}, \\
a^d_{\text{sl}} &=& (1.25 \pm 0.87) \% ~\text{($\mu D^*$ channel)}.
\end{eqnarray}
These values are consistent with those from the full analysis method. The uncertainties are larger as a result of the additional parameter in the fit.
The constant asymmetry parameter converges to values consistent with zero for both channels:
\begin{eqnarray}
A_{\text{const}} &=& (-0.03 \pm 0.23) \% ~\text{($\mu D$ channel)}, \\
A_{\text{const}} &=& (-0.09 \pm 0.21) \% ~\text{($\mu D^*$ channel)},
\end{eqnarray}
demonstrating that any possible residual background asymmetries not accounted for are small, as expected.

\section{\label{sec:comb}Combinations with Other Measurements}

\begin{figure*}[ht]
        \centering
        \subfigure[~Using combination of D0 and $B$ factory average for $a^d_{\text{sl}}$.]
        {\label{fig:combo1}\includegraphics[width=\columnwidth]{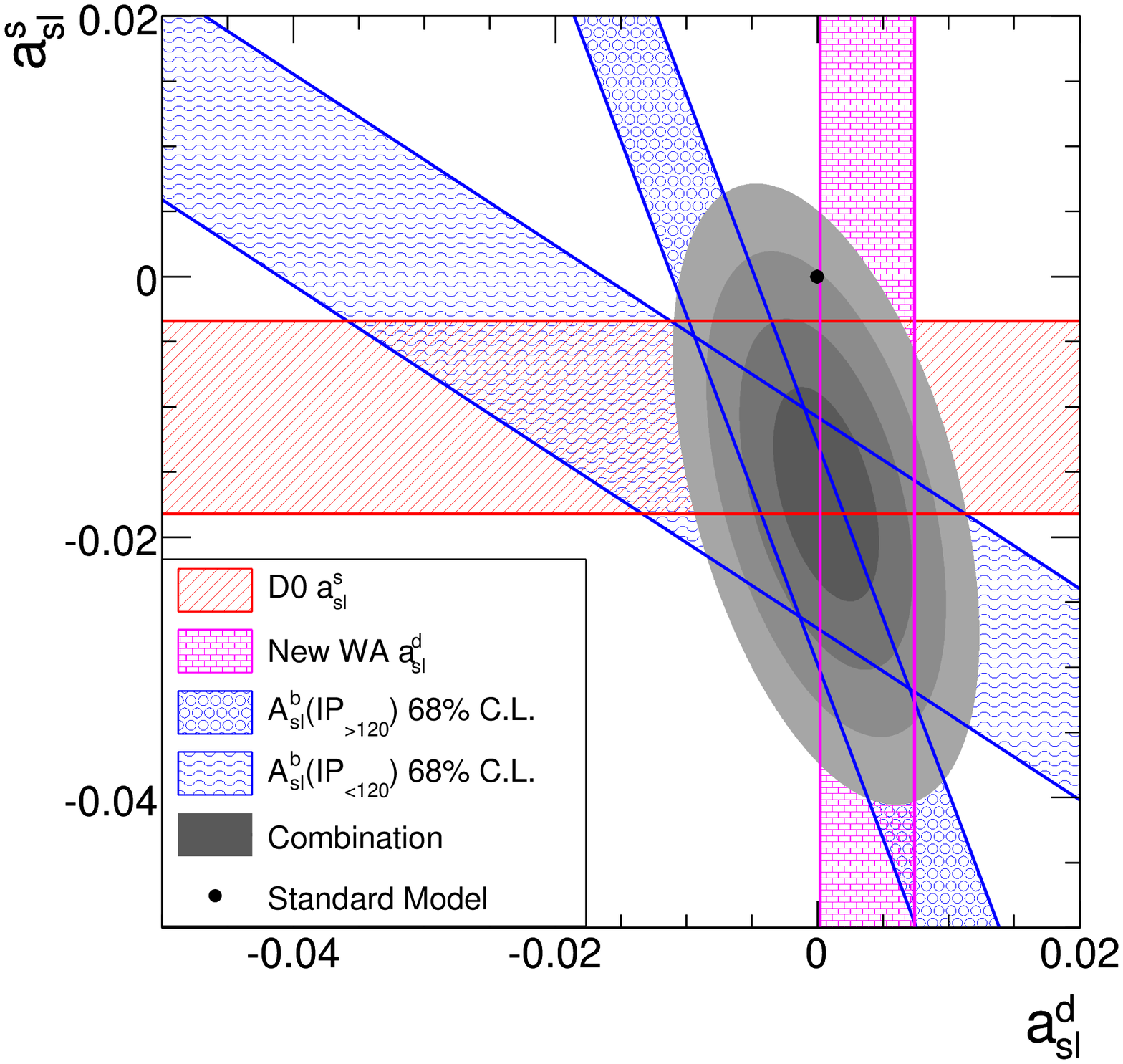}}
        \subfigure[~Using D0 value for $a^d_{\text{sl}}$.]
        {\label{fig:combo2}\includegraphics[width=\columnwidth]{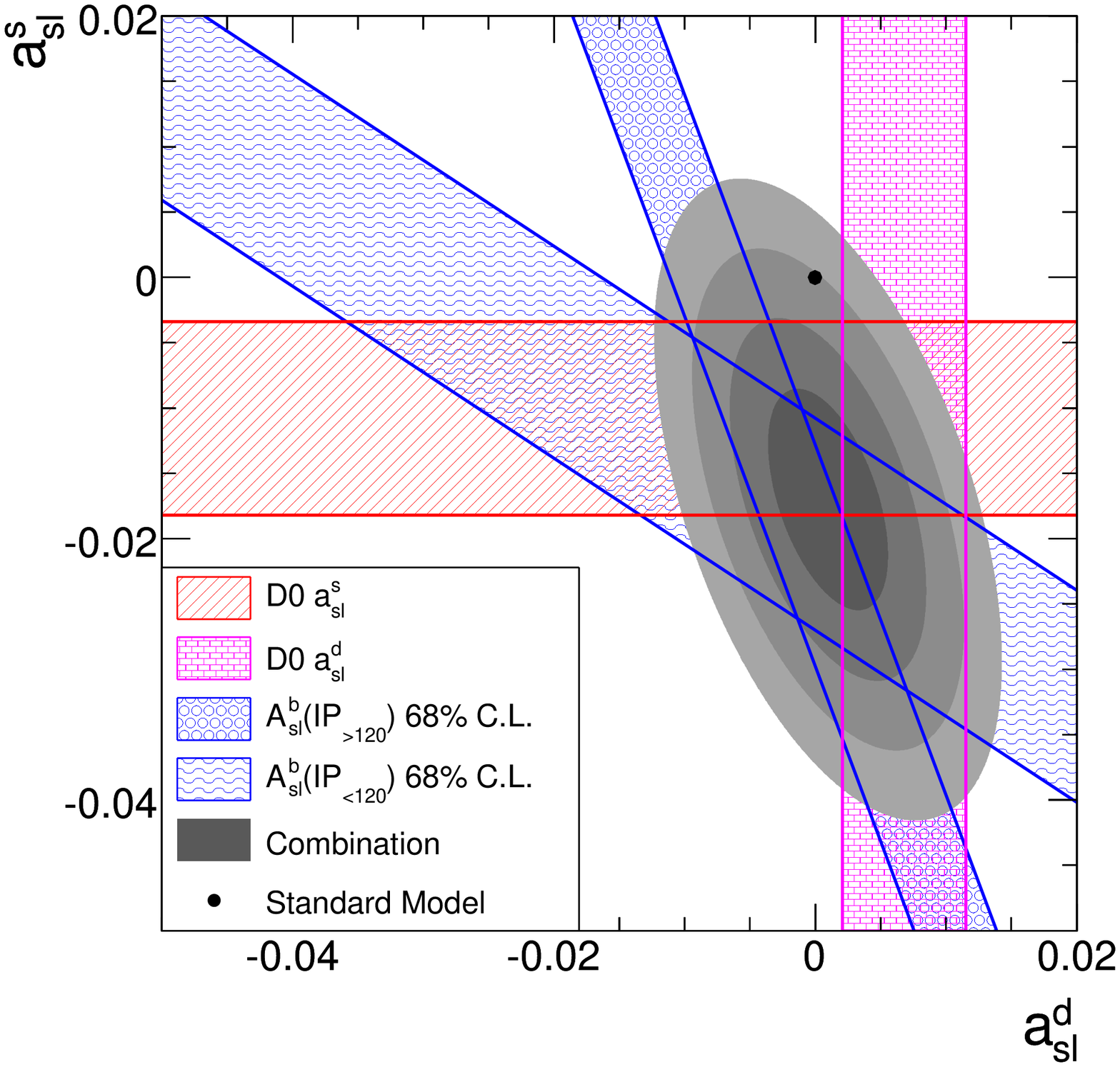}}
        \caption[]{Combination of measurements of $a^d_{\text{sl}}$ (D0 and existing world-average from B factories~\cite{pdg}), $a^s_{\text{sl}}$ (D0~\cite{assl_iain}), and the two impact-parameter-binned constraints from the same-charge dimuon asymmetry $A^b_{\text{sl}}$ (D0~\cite{d0_dimuon}). The bands represent the $\pm 1$ standard deviation uncertainties on each measurement. The ellipses represent the 1, 2, 3, and 4 standard deviation two-dimensional confidence level regions of the combination.}
\label{fig:combo}
\end{figure*}

This measurement of $a^d_{\text{sl}}$ can be combined with the existing world average from the $B$ factories~\cite{pdg}.
We use a simple weighted average, assuming that the two measurements are fully independent. The total uncertainty on the 
result presented in this article is $\pm 0.47$\%, obtained from the addition in quadrature of statistical and systematic 
uncertainties. The normalized weights are then $0.59$ (D0) and $0.41$ (WA).
We obtain:
\begin{eqnarray}
a^d_{\text{sl}} & = & (0.38 \pm 0.36) \%.
\end{eqnarray}
This number can in turn be combined with the recent $a^s_{\text{sl}}$ measurement~\cite{assl_iain}, and the two-dimensional 
constraints on ($a^d_{\text{sl}}$, $a^s_{\text{sl}}$) from the D0 measurement of the dimuon charge asymmetry $A^b_{\text{sl}}$~\cite{d0_dimuon}.
The full two-dimensional fit yields the following values:
\begin{eqnarray}
a^d_{\text{sl}}(\text{comb.}) & = & (0.07 \pm 0.27)\%, \\
a^s_{\text{sl}}(\text{comb.}) & = & (-1.67 \pm 0.54)\%, 
\end{eqnarray}
where the two parameters have a correlation coefficient of $-0.46$. The results are shown in Fig.~\ref{fig:combo1}, with the 
two dimensional contours overlaid on the four constraints from the input measurements. 
The fit returns a $\chi^2$ of 2.0 for 2 degrees-of-freedom.
The $p$-value of the combination with respect to the SM point is $0.0037$, 
corresponding to an inconsistency at the 2.9 standard deviation level.

Using only the D0 measurements of $a^d_{\text{sl}}$, $a^s_{\text{sl}}$, and $A^b_{\text{sl}}$, we obtain the 
following values:
\begin{eqnarray}
a^d_{\text{sl}}(\text{comb.}) & = & (0.10 \pm 0.30)\%, \\
a^s_{\text{sl}}(\text{comb.}) & = & (-1.70 \pm 0.56)\%, 
\end{eqnarray}
with a correlation coefficient of $-0.50$. The $\chi^2$ of this fit is 2.9, and the standard model $p$-value is $0.0036$,
corresponding to a 2.9 standard deviation effect. Figure~\ref{fig:combo2} shows the two-dimensional contours from this combination.

\section{\label{sec:conc}Conclusions}

We have performed a measurement of the semileptonic mixing asymmetry from $B^0$ decays, 
$a^d_{\text{sl}}$, using $B^0 \to \mu^+ D^{(*)-}X $ decays in two independent channels. 
We obtain $a^d_{\text{sl}} = [0.68 \pm 0.45 \text{ (stat.)} \pm 0.14 \text{ (syst.)}]\%$,
which is consistent with the SM prediction of $(-0.041 \pm 0.006) \%$.
The resulting precision is dominated by limited statistics in the signal channel, and is better 
than the current world-average precision obtained by combining results from the $B$ factories [Eq.~(\ref{eq:adsl_pdg})]. 
 
The background asymmetries are determined using data-driven methods, in dedicated decay channels. 
The most important background is from differences in the reconstruction efficiencies for 
positively and negatively charged kaons, which is of order $1\%$.
The use of simulation is limited to measuring the relatively small ($\sim$10--20\%) fraction 
of signal events which do not arise from $B^0$ decay, and modeling the oscillation of $B^0$ mesons.


%
We thank the staffs at Fermilab and collaborating institutions,
and acknowledge support from the
DOE and NSF (USA);
CEA and CNRS/IN2P3 (France);
MON, Rosatom and RFBR (Russia);
CNPq, FAPERJ, FAPESP and FUNDUNESP (Brazil);
DAE and DST (India);
Colciencias (Colombia);
CONACyT (Mexico);
NRF (Korea);
FOM (The Netherlands);
STFC and the Royal Society (United Kingdom);
MSMT and GACR (Czech Republic);
BMBF and DFG (Germany);
SFI (Ireland);
The Swedish Research Council (Sweden);
and
CAS and CNSF (China).
%

\end{document}